\documentclass[a4paper, 10pt, conference]{IEEEtran}
\IEEEoverridecommandlockouts
\usepackage{cite}
\usepackage{amsmath,amssymb,amsfonts}
\usepackage{graphicx}
\usepackage{textcomp}
\usepackage{soul}
\usepackage{color}
\usepackage{booktabs}
\usepackage{algorithm}
\usepackage{algorithmic}
\renewcommand{\algorithmicrequire}{\textbf{Input:}}  
\renewcommand{\algorithmicensure}{\textbf{Output:}} 
\newcommand{\tr}{\rm{Tr}}

\usepackage{multirow}

\usepackage{xcolor}
\def\BibTeX{{\rm B\kern-.05em{\sc i\kern-.025em b}\kern-.08em
    T\kern-.1667em\lower.7ex\hbox{E}\kern-.125emX}}

\begin{document}

\title{Fast State Stabilization using Deep Reinforcement Learning for Measurement-based Quantum Feedback Control}

\author
{Chunxiang Song, Yanan Liu, Daoyi Dong, Hidehiro Yonezawa
\thanks{Chunxiang Song is with the School of Engineering and Technology, University of New South Wales, Canberra, ACT 2600, Australia {(\tt\small chunxsong@gmail.com).}}
\thanks{Yanan Liu is with the School of Engineering, University of Newcastle, Callaghan, NSW 2308, Australia {(\tt\small yaananliu@gmail.com).}}
\thanks{Daoyi Dong is with the School of Engineering, Australian National University, Canberra, ACT 2601, Australia {(\tt\small daoyidong@gmail.com).}}
\thanks{Hidehiro Yonezawa is with the Optical Quantum Control Research Team, RIKEN Center for Quantum Computing, 2-1 Hirosawa, Wako, Saitama, 351-0198, Japan {(\tt\small hidehiro.yonezawa@riken.jp).}}
}

\maketitle

\begin{abstract}
\noindent The stabilization of quantum states is a fundamental problem for realizing various quantum technologies. Measurement-based-feedback strategies have demonstrated powerful performance, and the construction of quantum control signals using measurement information has attracted great interest. However, the interaction between quantum systems and the environment is inevitable, especially when measurements are introduced, which leads to decoherence. To mitigate decoherence, it is desirable to stabilize quantum systems faster, thereby reducing the time of interaction with the environment. In this paper, we utilize information obtained from measurement and apply deep reinforcement learning (DRL) algorithms, without explicitly constructing specific complex measurement-control mappings, to rapidly drive random initial quantum state to the target state. The proposed DRL algorithm has the ability to speed up the convergence to a target state, which shortens the interaction between quantum systems and their environments to protect coherence. Simulations are performed on two-qubit and three-qubit systems, and the results show that our algorithm can successfully stabilize random initial quantum system to the target entangled state, with a convergence time faster than traditional methods such as Lyapunov feedback control  and several DRL algorithms with different reward functions.  Moreover, it exhibits robustness against imperfect measurements and delays in system evolution.
\end{abstract}


\section{Introduction}
\noindent Quantum control theory focuses on manipulating quantum systems using external control fields or operations to regulate their behaviors \cite{dong2023learning}. A significant objective in quantum control is the preparation of target states, particularly quantum entangled states, which serve as vital resources for various quantum applications, including quantum teleportation \cite{karlsson1998quantum,ekert1998quantum}, fast quantum algorithms \cite{ekert1998quantum,jozsa2003role}, and quantum computations \cite{gu2019quantum}. Achieving high-fidelity entangled states often involves in using classical control methods, with feedback control technology being particularly noteworthy. Quantum systems can be stabilized at target states or spaces through feedback control methods that continuously monitor the system and design feedback controllers based on real-time feedback information. In quantum measurement-based feedback, quantum measurements, while providing valuable information, introduce stochastic noise, complicating the state preparation process. To address the challenges posed by stochastic nonlinear problems in quantum systems due to measurements, some classical control methods, such as the Lyapunov method \cite{kuang2017rapid,liu2021two,liu2016lyapunov}, have been applied. However, devising feedback strategies remains a formidable task, given the vast space of possibilities where different responses may be required for each measurement outcome. Moreover, opportunities exist for further enhancing stability and convergence speed. 

Recently, quantum learning control, first introduced in \cite{judson1992teaching}, has proven potential in addressing various quantum control problems. Its popularity has grown with the incorporation of additional machine learning (ML) algorithms that exhibit excellent optimization performance and promising outcomes. Quantum learning control concerns to apply proper ML algorithms as tools for improving quantum system performance \cite{dong2022quantum,judson1992teaching,Dong2020,sharma2010genetic,shir2008niching}, and can offer robust solutions for developing effective quantum control and estimation methods \cite{dong2022quantum}. For instance, the utilization of gradient algorithms in quantum learning control has demonstrated its significance in addressing quantum robust control problems \cite{dong2015sampling}. Another example involves a widely used class of algorithms, evolutionary algorithms (EAs), which gains attention in learning control due to their ability to avoid local optima and their independence from gradient information\cite{dong2023learning}. Nonetheless, the real-time nature and randomness of measurement feedback control pose challenges, expanding the decision space significantly due to randomness \cite{wiseman1994quantum}. This randomness makes it almost impossible to reproduce the same measurement trajectory, bringing challenges for EAs to apply control policies from certain sample trajectories to entirely different ones. 

Our study is motivated by the application of suitable deep reinforcement learning (DRL)  algorithms within feedback loops to exploit information obtained from measurements, thereby achieving predefined objectives. This approach holds the potential to enhance feedback control schemes, leading to a reduction in the convergence time to reach target states and exhibiting robustness in the face of uncertainties.  RL techniques have been applied for target state preparation in many-body quantum systems of interacting qubits, where a sequence of unitaries was applied and a measurement was implemented at the final time to provide reward information to the agent \cite{bukov2018reinforcement}. The similar idea was then utilized in harmonic oscillator systems, to achieve different target state preparations, through an ancilla qubit in \cite{sivak2022model}, where a final reward measurement (POVM) was carefully designed to provide reward information to the agent for different tasks of state preparation. In recent years, DRL (RL) approaches also started to play an important role in quantum algorithms, such as QAOA, for ground state preparation in different quantum systems \cite{wauters2020reinforcement,yao2021reinforcement}. The similar state preparation problem has also been considered in \cite{borah2021measurement}, where the system is a double-well system and the reward is a unique function of the measurement results. In \cite{porotti2022deep,perret2024preparation}, DRL-based approaches were employed for the preparation of cavity Fock state superpositions using fidelity-based reward functions, with system states as the training information under continuous weak measurement. In this study, we aim to devise a feedback control scheme based on DRL algorithms to enhance state stabilization, such as Bell states and GHZ states, for multi-qubit systems under continuous weak measurement. The designed algorithm can be applied to multi-qubit systems and provides a high fidelity and faster convergence to a given target state. 

To achieve the objectives, we exploit the information derived from quantum measurement as the input signal to train our DRL agent. The agent actively interacts with the quantum system, making control decisions based on the received input. We design a generalized reward function that quantifies the similarity between the current quantum state and the desired target state. This incentivizes the DRL agent to iteratively learn control strategies that maximize rewards, ultimately leading to more effective control strategies for stabilizing entangled quantum states. Our work shows the potential of DRL in solving complex quantum control challenges, contributing to the fields of quantum information processing and quantum computation. 

The main contributions of this paper are as follows:
\begin{enumerate}{}{}
    \item A DRL algorithm is proposed to achieve the stabilization of given entangled states in multi-qubit systems under continuous measurement. We design an effective and versatile reward function based on the distance between the system state and the target state, allowing flexible parameter adjustment for different objectives to enhance the performance of the DRL agent.
    \item We compare the proposed DRL-based control strategy with the Lyapunov method and other DRL methods, for state preparation through numerical simulations. Our DRL method achieves a faster stabilization for both target states, which effectively reduces the noise generated during system-environment interactions.
    \item We analyze the robustness of our DRL scheme under the presence of imperfect measurements and time delays in the feedback loop. The trained DRL agent exhibits remarkable adaptability to uncertainties in the environment, particularly excelling in the pursuit of robust control fields to achieve quantum state stability.
    \item {We conduct ablation studies to showcase the superiority of the proposed reward function. By comparing several commonly used reward function designs, our reward function shows better performance in stabilizing the target states.}
\end{enumerate}

The following is the organization of this paper. Section \ref{problemformulation} briefly introduces the stochastic master equation for quantum systems under continuous weak measurements. Section \ref{DRL}  explains in detail the logic and implementation behind DRL. Section \ref{application} gives some details of the implementation of DRL in the quantum measurement feedback control. Numerical results are given in Section \ref{Numrical}. In Section \ref{sec:CompRewards}, the performance of the proposed algorithm is analyzed through ablation studies and comparisons with other related methods. Section \ref{Conclusion} is the conclusion.

\section{Quantum System Dynamics}
\label{problemformulation}
\noindent For a quantum system, its state can be represented by a density matrix $\rho$ defined in the Hilbert space $\mathbb{H}$. This density matrix exhibits essential properties: Hermitian ($\rho = \rho^\dagger$), trace unity (${\tr}(\rho) = 1$), and positive semi-definite ($\rho \geq 0$). The dynamics of the quantum system may be observed through continuous weak measurements, enabling us to acquire pertinent measurement information for the design of an appropriate feedback controller. The evolution of the quantum trajectory can be described by the stochastic master equation (SME) \cite{wiseman1994quantum, doherty2000quantum}:
\begin{equation}
    \begin{aligned}
    d\rho_{t} = & -\frac{i}{\hbar}[H_0 + \sum_{j=1}^M u_j (t) H_j,\rho_{t}]dt \\
& + \kappa_c\mathcal{D}[c]\rho_{t}dt + \sqrt{\eta_c\kappa_c} \mathcal{H}[c]\rho_{t}dW, 
    \end{aligned}
\label{eq:SME}
\end{equation}
where $i=\sqrt{-1}$, the reduced Planck constant $\hbar=1$ is used throughout this paper; the Hermitian operator $H_0$ and $H_j$ $(j=1,2,\cdots,M)$ are the free Hamiltonian and control Hamiltonians, respectively; $u_j(t)$ is a real-valued control signal, which can be interpreted as the strength of the corresponding control Hamiltonians $H_j$; {$\kappa_c$ and $\eta_c$ are measurement strength and efficiency, respectively;} $dW$ is a standard Wiener process caused by measurement; the Hermitian operator $c$ is an observable; the superoperators $\mathcal{D}[c]\rho_{t}$ and $\mathcal{H}[c]\rho_{t}$ are related to the measurement, e.g., they can describe the disturbance to the system state, and the information gain from the measurement process, respectively \cite{jacobs2006straightforward}, and have the following forms:
\begin{equation}\label{dchc}
\begin{cases}
\mathcal{D}[c]\rho_{t}=c\rho_{t} c^\dagger -\frac{1}{2}c^\dagger c\rho_{t} -\frac{1}{2}\rho_{t} c^\dagger c,
\\
\mathcal{H}[c]\rho_{t}=c\rho_{t} + \rho_{t} c^\dagger - \tr[(c+c^\dagger)\rho_{t}]\rho_{t}.
\end{cases}
\end{equation}
On any given trajectory the corresponding measured current is $I(t) = dy(t)/dt$ \cite{wiseman2009quantum,doherty2000quantum} where 
\begin{equation}\label{dyt}
{dy(t) = \sqrt{\eta_c\kappa_c}\tr[(c+c^\dagger)\rho_{t}]dt + dW.
}\end{equation}
With the measurement result $y_t$, information on the standard Wiener process $dW$ can be collected from \eqref{dyt}. Utilizing \eqref{eq:SME}, an estimate of the system state can be obtained and utilized to construct a feedback controller.

In this paper, we consider the DRL-based feedback stabilization of the target quantum states. We will show our algorithm in stabilizing a GHZ entangled states of a three-qubit system and an eigenstate of an angular momentum system, while our scheme has the potential to be extended to other quantum systems.

\section{Deep Reinforcement Learning}
\label{DRL}
\noindent Abstracting a real-world problem into a Markov decision process (MDP) serves as the foundational step in applying DRL \cite{van2012reinforcement}. MDP provides a formal framework for modeling the interaction between an agent and its environment (quantum systems in this work), offering a structured specification of the agent's decision-making problem. The environment is abstracted with essential elements such as states, actions, rewards, and more. The agent is a pivotal component of DRL, representing a learning entity or decision-maker that, through interactions with the environment, learns to take actions to achieve objectives and continually refines its decision strategies to enhance the effectiveness of its actions. This process of agent-environment interaction and learning constitutes the core mechanism through which DRL efficiently tackles real-world challenges and achieves desirable outcomes.

An MDP is a structured representation denoted by the tuple $<\mathcal{S},\mathcal{A},R,P,\gamma>$, where each element serves as a crucial role in modeling the problem and applying DRL:
\begin{itemize}
    \item $\mathcal{S}=\{\rho \in \mathbb{H}:\rho=\rho^\dagger,\rm Tr(\rho)=1, \rho\ge0\}$ represents the set of states. At each time step $t$, the environment presents a specific quantum state $\rho_t$ to the agent, who subsequently makes decisions based on this state.
    \item $\mathcal{A}$ signifies the set of actions, incorporating the actions $a_t \in \mathcal{A}$ that the agent can undertake at each time step. In this context, the actions correspond to the control signals $u_j(t)$ defined in \eqref{eq:SME}, with values ranging from any bounded control strength, for example, $u_j(t)\in[-1,1]$ in this paper.
    \item $R$ denotes the reward function. This paper considers the task of stabilizing the current state to the target state, thus the immediate reward $r_t$ can be simplified as
    \begin{equation} \label{imreward}
        r_t = R(\rho_t).
    \end{equation}
    In this study, the reward function is defined using the trace-based distance $D_{\rho_t}$:
    \begin{equation} \label{distanceeq}
        D_{\rho_t} \rm Triangleq 1 - \rm Tr(\rho_d\rho_t), 
    \end{equation}
    which quantifies the difference between the current state $\rho_t$ and the target state $\rho_d$. When $D_{\rho_t} = 0$, the system state has stabilized at the target state $\rho_d$. 
    \item $P(\rho_{t+1}|\rho_{t},a_{t})$ is the state transition function. It indicates how the environment transitions to the next state $\rho_{t+1}$ after taking action $a_t$ in the current state $\rho_t$. It is consistent with the stochastic evolution of the quantum system described in \eqref{eq:SME}.
    \item $\gamma$ is the discount factor, which determines the emphasis placed on future rewards, influencing the agent's decision-making process in a long-term perspective.
\end{itemize}

\begin{figure}[!t]
\centerline{\includegraphics[width=0.49\textwidth]{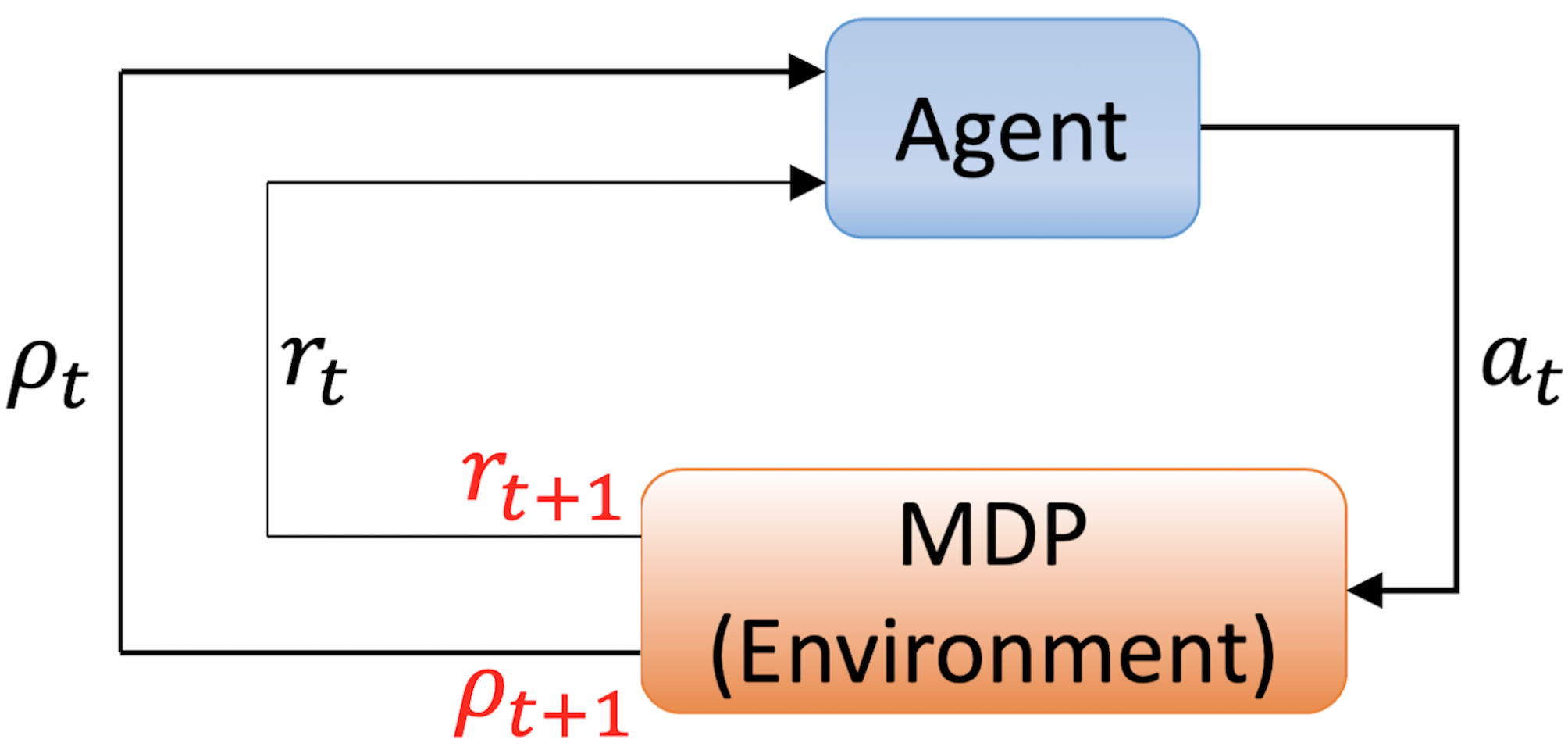}}
\caption{Illustration of MDP for Agent-Environment interaction.}
\label{MDP}
\end{figure}

An MDP is a time-dependent and ongoing process, involving continuous interaction between the agent and the environment. The interaction can be illustrated as shown in Fig.~\ref{MDP}. {For simplicity, we consider the interaction between the agent and the environment as a discrete time series (e.g., $t = 0, 1, 2, 3, \cdots$). Starting from $t=0$ with the known initial state $\rho_0$, the agent selects an action $a_0$ based on $\rho_0$ and applies it to the environment. Subsequently, the environment undergoes a state transition according to the state transition function $P(\rho_{t+1}|\rho_{t},a_{t})$, resulting in the next state $\rho_{1}$, and provides an immediate reward $r_1$ based on the reward function $R$. The environment also utilizes a classical computer to solve the SME \eqref{eq:SME} to estimate the density matrix of $\rho_{1}$, which is then fed back to the agent. This process is iterated until completion. Therefore, the MDP and the agent jointly generate a sequence or trajectory as follows:
\begin{equation}
\{\rho_0,r_0,a_0\},\{\rho_1,r_1,a_1\},\{\rho_2,r_2,a_2\},\cdots
\end{equation}
}
The function that selects an action $a$ from the set of actions $\mathcal{A}$ based on the current state $\rho$ is referred to as a policy $\pi(a|\rho)$. The objective of the MDP is to find the policy $\pi(a|\rho)$ that allows the agent to make optimal decisions, effectively maximizing long-term cumulative rewards. Different methods, including value-based, policy-based and actor-critic-based, have been developed to update the policy in the DRL \cite{mnih2013playing,sutton1999policy,konda1999actor}. In this paper, a highly effective actor-critic style proximal policy optimization (PPO) algorithm \cite{schulman2017proximal} will be applied. 

\section{Applying DRL to Quantum Measurement-based Feedback Control}
\label{application}
\noindent In this section, we apply the DRL to the quantum systems and aim to design a measurement-based feedback strategy to stabilize a given target state. The application is comprised of training and testing parts. In the training part, the primary objective lies in  the agent's policy function  $\pi_\theta(a|\rho)$, constructed by a neural network with the adjustable parameter set $\theta$. This parameter set is updated aiming for a higher reward by using data that is generated through the interaction between the agent and the environment. Once the agent finishes training, it can be applied to the quantum systems to generate real-time feedback control signals in achieving the stabilization of target states.

\subsection{Environment: States and Actions}
\noindent The environment's state is represented by the quantum system's density matrix, $\rho$, which contains all the information about the quantum state. To use $\rho$ in DRL, it needs to be converted into a format suitable for neural networks. We achieve this by flattening the density matrix into a vector that includes both its real and imaginary parts. For example, for a single-qubit system with $\rho =\begin{bmatrix}\alpha_1+\beta_{1}i & \alpha_2+\beta_{2}i \\ \alpha_3+\beta_{3}i & \alpha_4+\beta_{4}i \end{bmatrix},$ it is converted to: $\rho :=\begin{bmatrix} \alpha_1,\alpha_2,\alpha_3,\alpha_4,\beta_1,\beta_2,\beta_3,\beta_4
\end{bmatrix}^T$.This process keeps all the quantum state information and makes it usable for the neural network.

The policy function $\pi_\theta(a|\rho)$ maps the quantum state's vectorized form to a set of control actions $a$. These actions correspond to the control signals applied to the control Hamiltonians $H_1, H_2, \dots, H_M$. For example, if there are two control Hamiltonians, the actions at each time step are given by $a:=\begin{bmatrix}u_1,u_2\end{bmatrix}^T$, where $u_1$ and $u_2$ are the control amplitudes.

The DRL algorithm trains the policy to select actions that stabilize the quantum state. At each time step $t$, the agent uses the policy to choose an action $a_t$ based on the state $\rho_t$. The environment then updates the quantum state to $\rho_{t+1}$ according to the system dynamics and provides a reward $r_t$. This reward encourages the agent to take actions that reduces the difference between $\rho_{t+1}$ and the target state $\rho_{\text{target}}$.

It is important to note that the resulting state $\rho$ may not be a valid quantum state due to the inherent randomness in measurements and cumulative errors in solving the SME using classical computers \cite{qi2013quantum}. {For example, the state matrix may contain non-physical (negative) eigenvalues. To address this issue, a suitable approach is to check the eigenvalues of the estimated matrix at each step. When negative values are encountered, the state should be projected back onto a valid physical state.} This can be achieved by finding the {closest density matrix} under the 2-norm \cite{smolin2012efficient}. This approximation ensures that a non-physical density matrix is transformed into the most probable positive semi-definite quantum state with a trace equal to $1$.

As mentioned in Section \ref{DRL}, the interaction between the agent and the environment forms an episode $\tau = \{ \rho_0, r_0, a_0, \rho_1, r_1, a_1, \dots, \rho_m \},$ where $m$ is the final step of the episode, $\tau \in \mathbb{D} = \{\tau_n\}_{n=1,2,\dots,N}$, with $N$ denoting the total number of possible sequences. Without confusion, we use $\tau$ to represent a single episode. The primary goal of the agent is to maximize the expected cumulative reward across all possible episodes. The construction of the reward function will be discussed in detail in the following section.

\subsection{Reward}
\label{sec:myreward}
{\noindent The design of the reward function is critical in DRL. For instance, \cite{bukov2018reinforcement,sivak2022model} proposed sparse reward functions, i.e., providing rewards only at the end of each trajectory by evaluating the final state. While this approach is straightforward and easy to implement, it often suffers from slow learning or even training failure in complex systems due to the sparsity of reward signals. To address this, \cite{porotti2022deep} and \cite{perret2024preparation} introduced a scheme that collects fidelity information at each step and applies a higher-weighted reward for high-fidelity states. However, these methods still face challenges in balancing exploration and exploitation, particularly in systems with high-dimensional state spaces.}

In this work, we propose a {Partitioned Nonlinear Reward (PNR)} function based on the distance $D_{\rho_t}$ \eqref{distanceeq}. A lower $D_{\rho_t}$ indicates better alignment with the target state. The reward at each time step $r_t$ is defined as:
\begin{alignat}{1}
    r_t&=\left( \frac{\overline{D}-\underline{D}}{\mathfrak{f}*(D_{\rho_t}-\underline{D})-\mathfrak{e}*(D_{\rho_t}-\overline{D})}-\frac{1}{\mathfrak{f}}\right) \nonumber \\ 
    &\times \left(\frac{\mathfrak{e}\mathfrak{f}*(\overline{R}-\underline{R})}{\mathfrak{f}-\mathfrak{e}}\right)+\underline{R},\label{eq:rewardfunc} 
\end{alignat}
where $\overline{D}$ and $\underline{D}$ are the upper and lower bounds of the distance $D_{\rho_t}$, $\overline{R}$ and $\underline{R}$ are the upper and lower bounds of the reward, and $\mathfrak{e}$ and $\mathfrak{f}$ are parameters that regulate the slope of the reward curve. 
 
In general, this reward function is motivated from the inverse proportional function (the relation between $r_t$ and $D_{\rho_t}$ when other parameters in \eqref{eq:rewardfunc} are fixed), which ensures that the reward value increases when the distance $D_{\rho_t}$ decreases. More specifically, the bounds $\overline{D}$ and $\underline{D}$ are designed to ensure $D_{\rho_t}\in [\underline{D}, \overline{D}]$. The distance is mapped to a reward value constrained within the range $[\underline{R}, \overline{R}]$. The two tunable parameters, $\mathfrak{f}$ and $\mathfrak{e}$ are used to adjust the steepness when $D_{\rho_t}$ is approaching $\underline{D}$ and $\overline{D}$, respectively. 

\begin{figure}[!t]
\centerline{\includegraphics[width=0.49\textwidth]{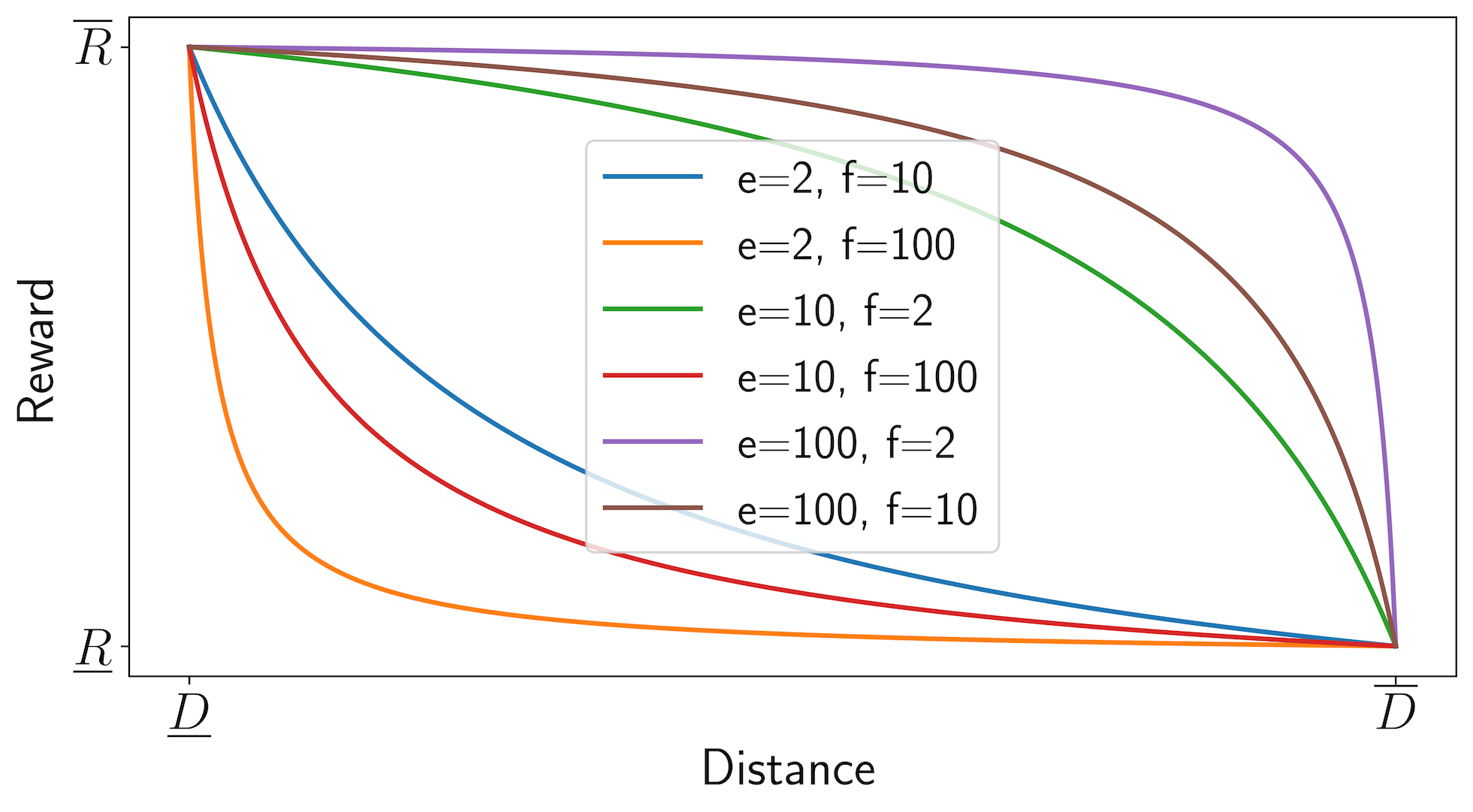}}
\caption{Example Reward Curves for Different Parameter Combinations of $\mathfrak{e}$ and $\mathfrak{f}$.}
\label{pic:rewardCurves}
\end{figure}

The detailed effects of different $\mathfrak{e}$ and $\mathfrak{f}$ are plotted in Fig.~\ref{pic:rewardCurves}. For the given bounds $\underline{D}, \overline{D}, \underline{R}$, and $ \overline{R}$, when $\mathfrak{e} < \mathfrak{f}$, the reward curve near $\underline{D}$ is steeper than that near $\overline{D}$. This indicates that the rate of increase in the reward value is more pronounced as $D_{\rho_t}$ approaches $\underline{D}$. Conversely, when $\mathfrak{e} > \mathfrak{f}$, the trend is reversed. 

In most existing work, the upper and lower bounds of $D_{\rho_t}$ are fixed at $\overline{D}=1$ and $\underline{D}=0$, respectively. The reward function in Eq. (7), however, offers the flexibility to adjust these bounds, as well as the corresponding reward values $\overline{R}$ and $\underline{R}$. This flexibility enables the division of the state space into multiple regions, which is particularly useful for addressing complex problems. Moreover, by adjusting $\overline{D}$ and $\underline{D}$, it becomes possible to assign positive reward values in certain regions of the state space and negative reward values (penalties) in others. This feature allows for more efficient and effective optimization, resulting in improved performance.

\begin{figure}[!t]
\centerline{\includegraphics[width=0.49\textwidth]{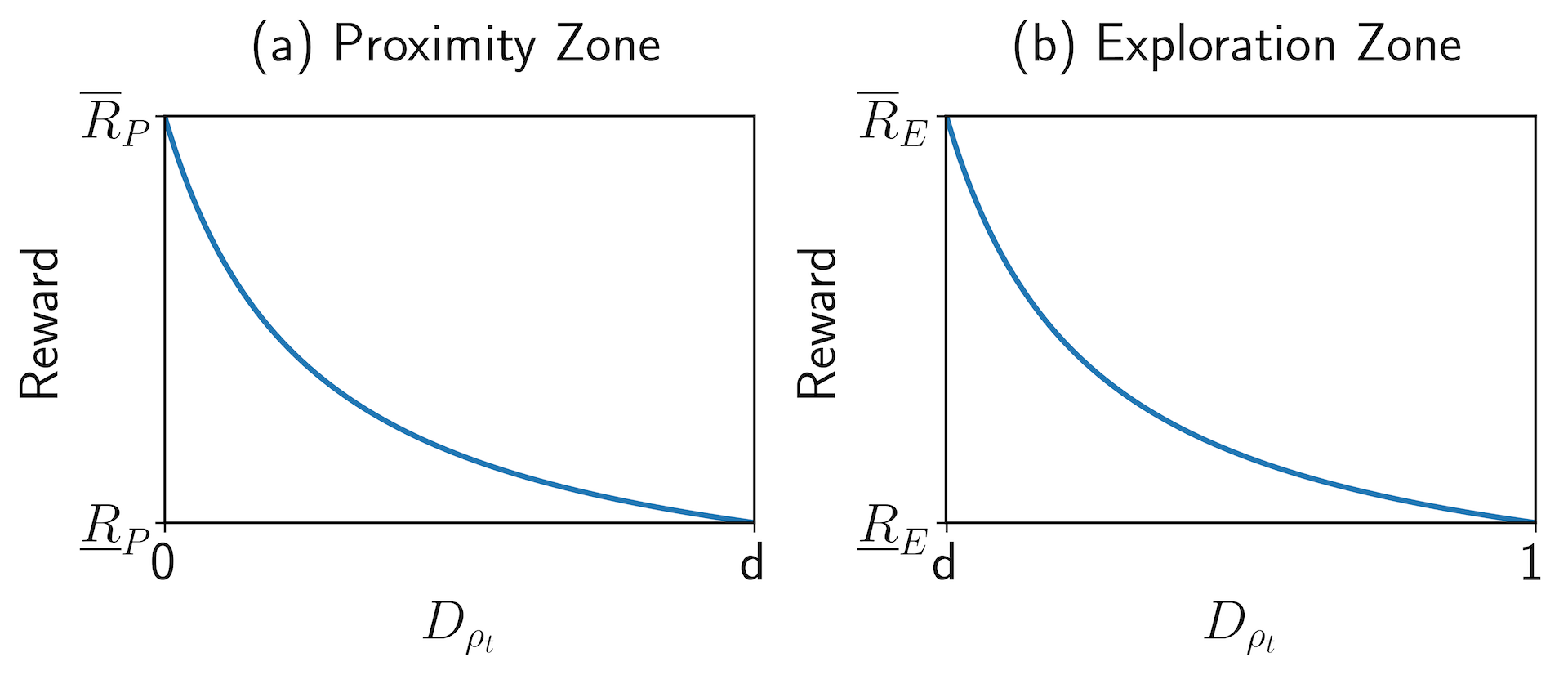}}
\caption{Reward Function. {(a) When the $D_{\rho_t}$ is between $0$ and $d$, we consider that the current system is almost approaching the target state. A smaller $D_{\rho_t}$ value corresponds to a larger positive reward. When $D_{\rho_t}=0$, the maximum reward $r = \overline{R}_P$ is obtained. (b) When the $D_{\rho_t}$ is between $d$ and $1$, we consider that there is still distance between the current state and the target state. At $D_{\rho_t}=1$, the maximum negative reward (penalty) is imposed, and as the $D_{\rho_t}$ value decreases, the penalty decreases accordingly.}}
\label{pic:rewardsZone}
\end{figure}

In this paper, we divide the state space into two regions using a partition parameter $d$. These two regions are denoted as \textit{Proximity Zone} ($D_{\rho_t} < d$) and \textit{Exploration Zone} ($D_{\rho_t} \geq d$). The same reward function in the formula \eqref{eq:rewardfunc} will be used in both of these two zones, and the corresponding bounds, $\underline{D}, \overline{D}$ are determined by the partition parameter $d$. This division provides more flexibility to balance the reward and the penalty in the whole state space. One possible curve of the reward function in these two zones are shown in Fig. ~\ref{pic:rewardsZone}.

In the \textit{Proximity Zone} ($D_{\rho_t} < d$), the state is considered close to success. Positive rewards encourage the agent to converge rapidly to the target state. The distance bounds in this region are $\overline{D}=\overline{D}_P = d$ and $\underline{D}=\underline{D}_P = 0$, while the reward bounds are $\overline{R}_P > \underline{R}_P \geq 0$. As shown in Fig.~\ref{pic:rewardsZone}(a), the reward function ensures that the slope of the curve increases as $D_{\rho_t}$ approaches zero, emphasizing rapid stabilization near the target state.

In the \textit{Exploration Zone} ($D_{\rho_t} \geq d$), the system is considered far from the target state, and penalties are applied to guide exploration. The boundaries of this zone are $\overline{D}=\overline{D}_E = 1$ and $\underline{D}=\underline{D}_E = d$, with the reward range satisfying $\underline{R}_E < \overline{R}_E \leq 0$. As illustrated in Fig.~\ref{pic:rewardsZone}(b), the penalty decreases as $D_{\rho_t}$ approaches $d$, with the slope of the curve ensuring a progressively faster reduction. This mechanism encourages the state to transition into the \textit{Proximity Zone}.

\textbf{Remark 1:} In the \textit{Exploration Zone}, when the distance $D_{\rho_t}$ approaches $d$, we reduce penalties rather than introduce large positive rewards. This approach avoids potential reward hacking \cite{hadfield2017inverse}, where the agent might exploit increasing rewards by staying near $D_{\rho_t} \approx d$ without fully reaching the target. By applying small penalties at each step, the agent is encouraged to explore more broadly while progressively improving its policy.

Additionally, a small penalty proportional to the number of steps taken before stabilization is introduced to encourage efficiency. For example, the first step incurs a penalty of $-1 \times 10^{-6}$, the second step $-2 \times 10^{-6}$, and so forth. {This step-based penalty discourages unnecessary delays and motivates the agent to achieve stabilization promptly.}

The proposed reward function adjusts based on the distance $D_{\rho_t}$, with steeper slopes in both regions as the system approaches the target state. This design encourages exploration in the \textit{Exploration Zone} while driving efficient convergence in the \textit{Proximity Zone}. {We show the superiority of our design by simulating different reward functions in Section \ref{sec:CompRewards}.}

With this reward function, we are now in the position to maximize the cumulative expected reward of the sequence, so for a complete sequences $\tau$, its cumulative reward can be expressed as
\begin{equation}
    \label{R_tau}
    R(\tau) = \sum_{t=0}^{m} A^{\theta}(\rho_t, a_t).
\end{equation}
$A^{\theta}(\rho_t, a_t) = Q(\rho_t, a_t) - V^{\phi}(\rho_t)$ is known as the advantage function in the field of RL, which is utilized to assess the desirability of taking a specific action $a_t$ at state $\rho_t$. $Q(\rho_t,a_t)=\sum_{t^{\prime}=t}^{m}\gamma^{t^{\prime}-t}r_{t^\prime}$ represents the action-value function, indicating the expected discounted reward for choosing action $a_t$ in state $\rho_t$, i.e., the cumulative sum of rewards until the end of the episode after executing this action. $r_{t^\prime}$ is the reward function in \eqref{eq:rewardfunc}. The value of $\gamma$ lies between $0$ and $1$, determining the emphasis on long-term rewards (close to $1$) or short-term rewards (close to $0$). It effectively introduces a discounting mechanism for future rewards, thereby shaping the agent's preference for future reward consideration when making decisions. $V^{\phi}(\rho_t)$ is referred to as the state-value function (or baseline) and is modeled by a neural network with the same structure as the policy network but with different parameters $\phi$. It is primarily employed to approximate the discounted rewards from state $\rho_t$ to the end of an episode. Specifically, if the current state is $\rho_t$, and for all possible actions $a_{t}^{(1)}, a_{t}^{(2)}, \dots, a_{t}^{(\dots)}$, they correspond to discounted rewards $Q(\rho_t, a_{t}^{(1)}), Q(\rho_t, a_{t}^{(2)}), \dots, Q(\rho_t, a_{t}^{(\dots)})$. As $V^{\phi}(\rho_t)$ represents the expected value of the discounted rewards at $\rho_t$, we can use $Q(\rho_t, a_{t}^{(1)}), Q(\rho_t, a_{t}^{(2)}), \dots, Q(\rho_t, a_{t}^{(\dots)})$ as features to approximate the value of $V^{\phi}(\rho_t)$, representing the expected value of rewards in state $\rho_t$. When $A^{\theta}(\rho_t, a_t) > 0$, action $a_t$ is considered better than average and is worth increasing the probability of being chosen in subsequent iterations while decreasing the probability otherwise.

\subsection{Training} \label{sec:training}
\noindent {The essence of employing DRL to address MDPs is centered on the training regimen of the agent, which is governed by the policy $\pi_{\theta}(a|\rho)$ . This study implements the PPO algorithm, a model-free policy gradient method within the domain of DRL. PPO incorporates a ``clip ratio'' mechanism that limits the extent of policy updates, thereby enhancing the stability and reliability of the learning process. This approach mitigates the requirement for extensive environmental sampling, which is a common feature of traditional policy gradient techniques, and consequently, it improves the efficiency of the training phase. The PPO algorithm operates with three sets of network parameters: the primary policy parameters, policy parameter duplicates, and value network parameters. These parameters are instrumental in the iterative policy refinement and state value estimation processes. The comprehensive algorithmic details and the underlying mathematical formalism are presented in the Appendix.}

Our DRL algorithm is implemented using the open-source Python library Stable-Baselines3 \cite{stable-baselines3}, while the quantum dynamic environment is constructed within the Gymnasium framework \cite{towers2024gymnasium}. {All simulations in this study are conducted on a computer equipped with an Apple M1 Pro chip and $32$ GB of memory, utilizing Python $3.11.3$, stable-baselines3 $2.3.2$, and Gymnasium $0.29.1$.} We design a reasonable reward function to guide the DRL agent through iterative learning, aiming to train an excellent DRL agent capable of generating control signals to achieve the stability of the target entangled state.

\noindent {{\bf{Reward Function Parameters:}} We set $d = 0.001$, which is the partition parameter of the reward function mentioned in Section \ref{sec:myreward}. This means that when the distance $D_{\rho_t}$ is less than $0.001$, the system is considered to be close to the target and receives a positive reward. In this case, the fidelity of the system state can easily exceed $0.999$. Ideally, the smaller the value of  $d$, the higher the control fidelity of the trained agent. However, as  $d$  decreases, the training process becomes increasingly challenging and time consuming. Therefore, the choice of $d$ should strike a balance between achieving high fidelity and managing training complexity.} 

{The upper and lower bounds of the reward in the \textit{Proximity Zone},  $\overline{R}_P$  and  $\underline{R}_P$ , are set to $100$ and $1$, respectively, encouraging the system to get as close as possible to the target. In the \textit{Exploration Zone}, the upper and lower bounds of  $\overline{R}_E$  and  $\underline{R}_E$  are set to $0$ and $-0.1$, respectively, maintaining a penalty without overly punishing the agent.}

We set $\mathfrak{e} = 2$ and $\mathfrak{f} = 10$, emphasizing a steeper reward escalation as $D_{\rho_t}$ approaches $\underline{D}$.

\noindent {\bf{Initial State:}} During training, the state is randomly reset to a quantum state after the completion of each episode, which means that at each episode in the training iteration, the agent starts from a new state and explores the environment from that point. 

\noindent {\bf{Early Termination:}} Regarding early termination, continuous quantum measurement feedback control can be modeled as an infinite-horizon MDP, but during training, each episode is simulated on a finite time horizon. Additionally, practical applications require a finite system evolution time. Therefore, we set fixed duration or termination conditions to end an episode. The termination conditions include the following:
\begin{itemize}
    \item If the system is measured continuously for $10$ iterations, and the distance  $D_{\rho_t}$  remains within the interval $[0, d]$ for all measurements, the system is considered to have {reached the target state with high fidelity}. At this point, the task is concluded.
    \item In a specific system, the maximum training time for a trajectory is set to a fixed value. For example, for the two-qubit state stabilization problem in Section \ref{2qubitSystem}, we set the maximum training time $T = 20$ arbitrary units (a.u.). When the evolution time reaches $20$ a.u., regardless of whether it has converged to the goal or not, the training trajectory is halted. This approach not only greatly saves training time but also significantly reduces the issue of overfitting. 
\end{itemize}
These early termination conditions bias the data distribution towards samples that are more relevant to the task, thereby saving training time and preventing undesirable behaviors. During agent testing, the time is typically not limited to evaluate the agent's performance in a real environment and assess its ability to complete tasks within a reasonable time frame.

The pseudo-code for the PPO in quantum state stabilization is shown in Algorithm \ref{alg1}.

\begin{algorithm}
	\renewcommand{\algorithmicrequire}{\textbf{Input:}}
	\renewcommand{\algorithmicensure}{\textbf{Output:}}
	\caption{PPO for Quantum Feedback Control}
	\label{alg1}
	\begin{algorithmic}
		\STATE \textbf{Begin}
		\STATE Initial policy weights $\theta$
        \STATE Initial value function weights $\phi$
		\REPEAT 
            \vspace{0.5em}
            \STATE Sampling training data from the environment
    		\FOR{t = 0,1,\dots}
    		\STATE $\rho_t \gets$  start state
            \STATE $r_{t} \gets$ reward
    		\STATE $a_t \thicksim \pi_{\theta}(a_t|\rho_t)$
    		\STATE Apply $a$ and simulate the system one step forward
    		\STATE $\rho_{t+1} \gets$  end state
            \STATE $r_{t+1} \gets$ reward
            \STATE Record $(\rho_t,r_t,a_t,\rho_{t+1})$ into memory $M$
            \ENDFOR
            
            \vspace{0.5em}
            \STATE $\theta^\prime \gets \theta$
            \vspace{0.5em}
            
            \FOR{each update}
                \STATE Sample $Nm$ samples $\{(\rho_t,r_t,a_t,\rho_{t+1})\}$ from $M$
                \vspace{0.5em}
                \STATE Update main policy:
                \FOR{each $(\rho_t,r_t,a_t,\rho_{t+1})$}
                    \STATE compute advantage $A^{\theta}$ in \eqref{R_tau} using current $V^{\phi}$ and GAE
                    \STATE compute $\frac{p_{\theta}(a_t | \rho_t)}{p_{\theta^{\prime}}(a_t|\rho_t)}$
                \ENDFOR
                \STATE Calculate the gradient to update the policy parameters $\theta$ via \eqref{PPO_Final}
                
                \vspace{0.5em}
                \STATE Update value function $V^{\phi}$:
                \FOR{each $(\rho_t,r_t,a_t,\rho_{t+1})$}
                    \STATE Use TD to update the value function $V^{\phi}$
                \ENDFOR
            \ENDFOR
		\UNTIL The training termination condition is triggered
	\end{algorithmic}  
\end{algorithm}

\section{Numerical Simulation}
\label{Numrical}
\subsection{Two-Qubit System}
\label{2qubitSystem}
\noindent We consider a two-qubit system in a symmetric dispersive interaction with an optical probe, as described in \cite{mirrahimi2007stabilizing}.  The system’s dynamics are governed by the SME in \eqref{eq:SME},  where we utilize a DRL control scheme to stabilize the system to a target entangled state from arbitrary initial states. Denote the Pauli matrices 
\begin{equation*}
    \sigma_x = \begin{bmatrix}0 & 1 \\ 1 & 0 \end{bmatrix}, \qquad \sigma_y = \begin{bmatrix}0 & -i \\ i & 0 \end{bmatrix}, \qquad \sigma_z = \begin{bmatrix}1 & 0 \\ 0 & -1 \end{bmatrix}. 
\end{equation*}
Control Hamiltonians in \eqref{eq:SME} are chosen as
\begin{equation}
H_1  = \sigma_y \otimes I_2, \quad \text{and} \quad   H_2 = I_2 \otimes \sigma_y,
\end{equation}

which implies that two control channels are applied to this quantum system. And the forms of $H_1$ and $H_2$ represent $\sigma_y$ rotations on the first and second qubits, respectively, enabling independent control over each qubit.

The observable operator is chosen as: 
\begin{equation}
    c  = \sigma_z \otimes I_2 + I_2 \otimes \sigma_z,
\end{equation}
which corresponds to the measurement of $\sigma_z$-like observable for the two qubits. 

Specify the target state as
\begin{equation}
    \rho_d = {\begin{bmatrix}0 & 0 & 0 & 0\\0 & 0.5 & 0.5 & 0 \\0 & 0.5 & 0.5 & 0 \\0 & 0 & 0 & 0 \end{bmatrix}},
\end{equation}
which is a symmetric two-qubit Bell state.

We utilize the previously summarized PPO algorithm to train the DRL agent. For the training trajectories, we set a time interval of $\Delta t = 0.001$ a.u. for each measurement step, with a maximum evolution time of $T=20$ a.u., corresponding to a maximum of $20,000$ steps. At each step, the DRL agent interacts with the environment, obtaining system information to generate control signals, which are then stored for iterative updates of the policy. The total number of training steps is set to $10^7$. {On the computer with configuration in Section \ref{sec:training}, the training process requires approximately $50$ minutes. However, the primary focus of this study is the performance of the trained agent rather than the optimization of training time. In practical implementations, employing a pre-trained agent is feasible. Once the target state is specified, the agent does not require repeated updates and can be directly utilized after training is completed.}

\begin{figure}[!t]
\centerline{\includegraphics[width=0.49\textwidth]{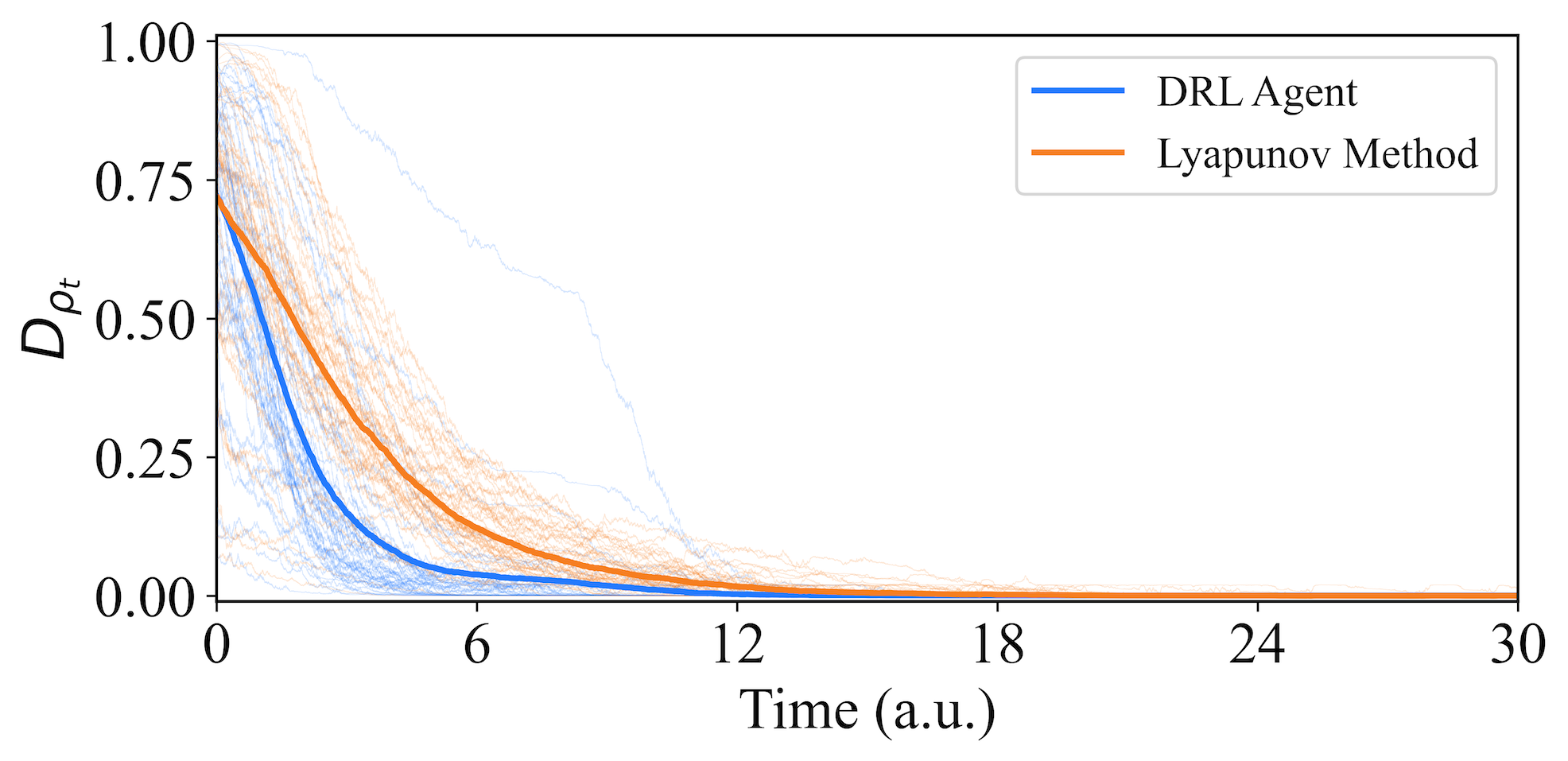}}
\caption{Evolution of the distance $D_{\rho_t}$ for $50$ random initial states stabilized to the target Two-Qubit state under the control of the DRL agent (blue) and Lyapunov method (orange). {The average stabilization time to the target under DRL control is $4.59$ a.u., while the Lyapunov method requires an average time of $5.86$ a.u.. (The stabilization time is defined as the time when the distance $D_{\rho_t} \leq 0.001$.)} The light blue (orange) lines represent the average evolution trajectories for different initial states, and the dark blue (orange) line represents the average trajectory across all different initial states.}
\label{pic:AM_100Samples}
\end{figure}

In order to evaluate its performance, we test the proposed strategy on $50$ randomly selected distinct initial quantum states $\rho_0$ (corresponding to different initial distances $D_{\rho_0}$, as indicated by the blue line in Fig.~\ref{pic:AM_100Samples}). {The light blue lines represent the evolution trajectory of a specific initial state with respect to the target state, averaged over $50$ different trajectories under varying environmental noise,} while the dark blue line depicts the average evolutionary trajectory of all the different initial states, {i.e., the average trajectory of $2500$ different evolutionary trajectories. Smaller values of the distance $D_{\rho_t}$ indicate that the system is closer to the target state. It is worth noting that a trained agent can successfully stabilize any initial state to the target state with high fidelity.} Furthermore, for comparison with the Lyapunov method mentioned in \cite{mirrahimi2007stabilizing}, we retain the same $50$ sets of randomly selected initial states and obtain the orange trajectories in Fig.~\ref{pic:AM_100Samples} using the Lyapunov method. It can be observed that the control signals generated by the DRL agent outperform the Lyapunov method. {Assuming that the time when the distance $D_{\rho_t}$ is less than $0.001$ is the evolution time of the system, the average evolution time of the $2500$ trajectories under the guidance of the DRL agent is $4.59$ a.u., while the average time using the Lyapunov method is $5.86$ a.u.. The DRL's average stabilization time is improved by $22\%$ over the Lyapunov method.} This indicates that our DRL approach successfully stabilizes these quantum states to the target state faster than the Lyapunov method.

\subsection{GHZ State}
\noindent We then consider a more complex problem of preparing three-qubit entangled GHZ states, which are special entangled states and have been regarded as maximally entangled states in many measures \cite{greenberger1989going}, \cite{dur2000three}. A GHZ entangled state is defined in the following form \cite{greenberger1989going,monz201114}:
{\begin{equation}\label{GHZ}
    |{\rm{GHZ}}\rangle = \frac{1}{\sqrt{2}} (|0\rangle^{\otimes n} + |1\rangle^{\otimes n}),
\end{equation}
where $n$ is the number of qubits.} Its density matrix can be expressed as $\rho_{\rm{GHZ}} \rm Triangleq |{\rm{GHZ}}\rangle \langle {\rm{GHZ}}|$. For the three-qubit GHZ state, we choose $|{\rm{GHZ}}\rangle = (1/\sqrt{2}) ({|000\rangle} + {|111\rangle})$, which gives the following  density matrix:
\begin{equation} \label{targetmatrix}
\begin{split}
    \rho_{\rm{GHZ}}  = \frac{1}{2} \begin{bmatrix} 1 & 0 & 0 & 0 & 0 & 0 & 0 & 1 \\ 0 & 0 & 0 & 0 & 0 & 0 & 0 & 0 \\ 0 & 0 & 0 & 0 & 0 & 0 & 0 & 0 \\ 0 & 0 & 0 & 0 & 0 & 0 & 0 & 0 \\ 0 & 0 & 0 & 0 & 0 & 0 & 0 & 0 \\ 0 & 0 & 0 & 0 & 0 & 0 & 0 & 0 \\ 0 & 0 & 0 & 0 & 0 & 0 & 0 & 0 \\ 1 & 0 & 0 & 0 & 0 & 0 & 0 & 1 \end{bmatrix}.    
\end{split}
\end{equation}
A degenerate observable $c$ is required according to the quantum state collapse after measurement \cite{mirrahimi2007stabilizing,liu2016lyapunov,kuang2021rapid}. The quantum state collapse states that the system in \eqref{eq:SME} will randomly converge to an eigenstate or eigenspace of $c$ without any control. Hence, we choose an observable in the following diagonal form:
\begin{equation} \label{measurementOperator}
    c = \text{diag} [\lambda_d,\lambda_2,\cdots,\lambda_{n-1},\lambda_d],
\end{equation}
where $\lambda_d \ne \lambda_k$ $(k=2,\cdots,n-1)$, and $\lambda_d$ is the eigenvalue corresponding to the target state $\rho_d$, i.e., $c\rho_d=\lambda_d \rho_d$.

Due to the degenerate form of the observable $c$ in \eqref{measurementOperator}, the system may converge to other state in the corresponding eigenspace related to $\lambda_d$, two control channels $H_1$ and $H_2$ based on Lyapunov method have been applied in \cite{mirrahimi2007stabilizing,liu2016lyapunov,kuang2021rapid} to solve this problem. For subsequent performance comparisons, two control channels are also used in this paper.

For any training trajectory, we take a time interval $\Delta t = 0.001$ a.u. for each measurement step. Given that we have set the maximum evolution time as $T=40$ a.u., it means the maximum number of evolution steps for any trajectory during training is $40,000$. The total number of training steps is $10^8$.

For all instances, in order to compare with the Lyapunov methods presented in \cite{liu2016lyapunov}, we choose the same system Hamiltonian as $H_0=\text{diag}[1,-1,-1,1,1,-1,-1,1]$.  The diagonal form of which indicates that the eigenvalues correspond to the energy levels of the system in the computational basis $|q_1q_2q_3\rangle$ (where $q_i\in \{0,1\}$ for three qubits).  The target state is $\rho_d \rm Triangleq \rho_{\rm{GHZ}}$ \eqref{targetmatrix}, and the observable $c$ is chosen to be:
\begin{alignat}{1}
    c & = 2 \times (\sigma_z \otimes \sigma_z \otimes I_2) + I_2 \otimes \sigma_z \otimes \sigma_z \nonumber \\
    & = \text{diag}[3,1,-3,-1,-1,-3,1,3],
\end{alignat}
 to measure correlations between the z-components of different pairs of qubits in a three-qubit system and also ensure that the target state is an eigenstate. 
The control Hamiltonians $H_1$ and $H_2$ are chosen as
\begin{equation}
    H_1 = I_2 \otimes I_2 \otimes \sigma_x + \sigma_x \otimes \sigma_x \otimes I_2, 
\end{equation}
and
\begin{equation}
    H_2 = \sigma_x \otimes I_2 \otimes I_2 + I_2 \otimes \sigma_x \otimes \sigma_x.
\end{equation}
 Similar to the two-qubit systems, two control channels represented by $H_1$ and $H_2$ together with their strengths $u_1$  and  $u_2$ provide control over the system’s dynamics. The forms of control Hamiltonian represent how the control action is applied to the three-qubit system. For example, the form of $H_1$ generally represents an independent $x$-axis control to the third qubit and a correlated $x$-axis interaction between the first and the second qubits.

We test the trained DRL agent in various environments to evaluate its performance and robustness. The goal is to assess how well the agent generalizes its learned policies to different scenarios and how it copes with perturbations and variations in the environment. To achieve this, we expose the trained DRL agent to a set of diverse environments, each with unique characteristics and challenges. These environments are carefully designed to represent a wide range of scenarios and potential disturbances that the agent might encounter in real-world applications. During the testing phase, we measure the agent's performance in terms of its ability to achieve the desired objectives and maintain stability in each environment. In particular, we examine its response to changes in the measurement efficiency $\eta_c$ and time delay disturbances to assess its robustness and adaptability.

We first investigate the ``perfect case''. In this paper, the ``perfect case'' entails assuming that negligible delay in solving the SME \eqref{eq:SME} by classical computers, and perfect detection, that is, measurement efficiency $\eta_c=1$. In contrast, situations where there is delay or imperfect detection within the system are collectively referred to as ``imperfect cases''. We then show some performance indications for ``imperfect cases''.

\subsubsection{Stabilization of the GHZ state under perfect case}

\begin{figure}[!t]
\centerline{\includegraphics[width=0.49\textwidth]{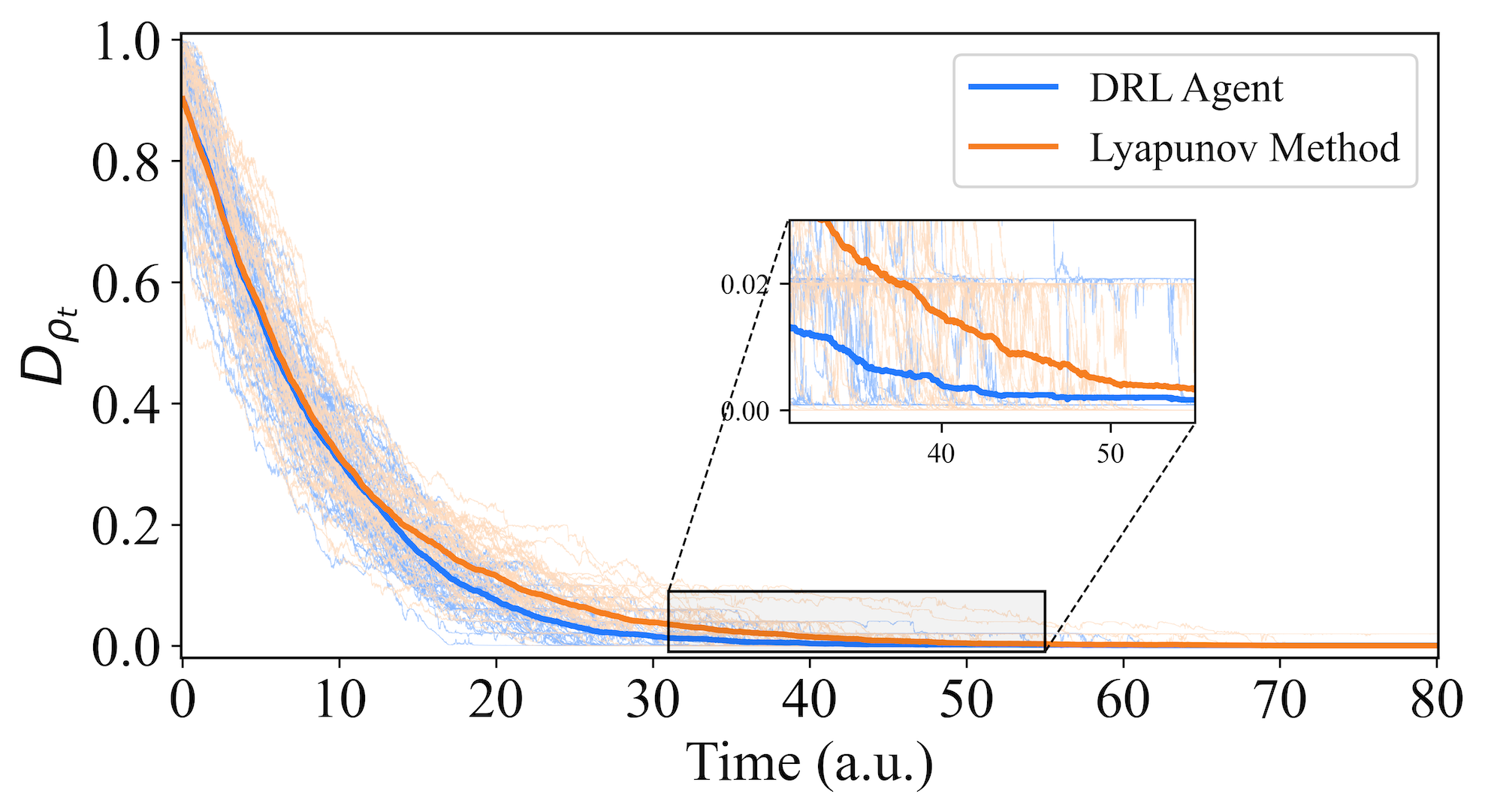}}
\caption{Evolution of the distance $D_{\rho_t}$ for $50$ random initial states stabilized to the target GHZ state under the control of the DRL agent (blue) and Lyapunov method (orange). {The average stabilization time to the target under DRL control is $10.41$ a.u., while the Lyapunov method requires an average time of $12.33$ a.u.. (The stabilization time is defined as the time when the distance $D_{\rho_t} \leq 0.001$.)} The light blue (orange) lines represent the average evolution trajectories for different initial states, and the dark blue (orange) line represents the average trajectory across all different initial states.}
\label{GHZ_100Samples}
\end{figure}

We initiate the testing phase to evaluate the ability of arbitrary initial states to stabilize to the target GHZ state within a specified time frame. As shown in Fig.~\ref{GHZ_100Samples}, we employ the comparative approach mentioned in Section \ref{2qubitSystem}, randomly selecting $50$ distinct initial states for control using the DRL agent and the Lyapunov method. The blue and orange lines correspond to the DRL method and the Lyapunov method, respectively. {The average evolution time of 2500 trajectories guided by the DRL agent is $16\%$ shorter than that of the Lyapunov method (DRL: 10.41 a.u. vs. Lyapunov: 12.33 a.u.).} This indicates that our DRL approach successfully stabilizes quantum states to the target GHZ state more rapidly.

\begin{figure}[!t]
\centerline{\includegraphics[width=0.49\textwidth]{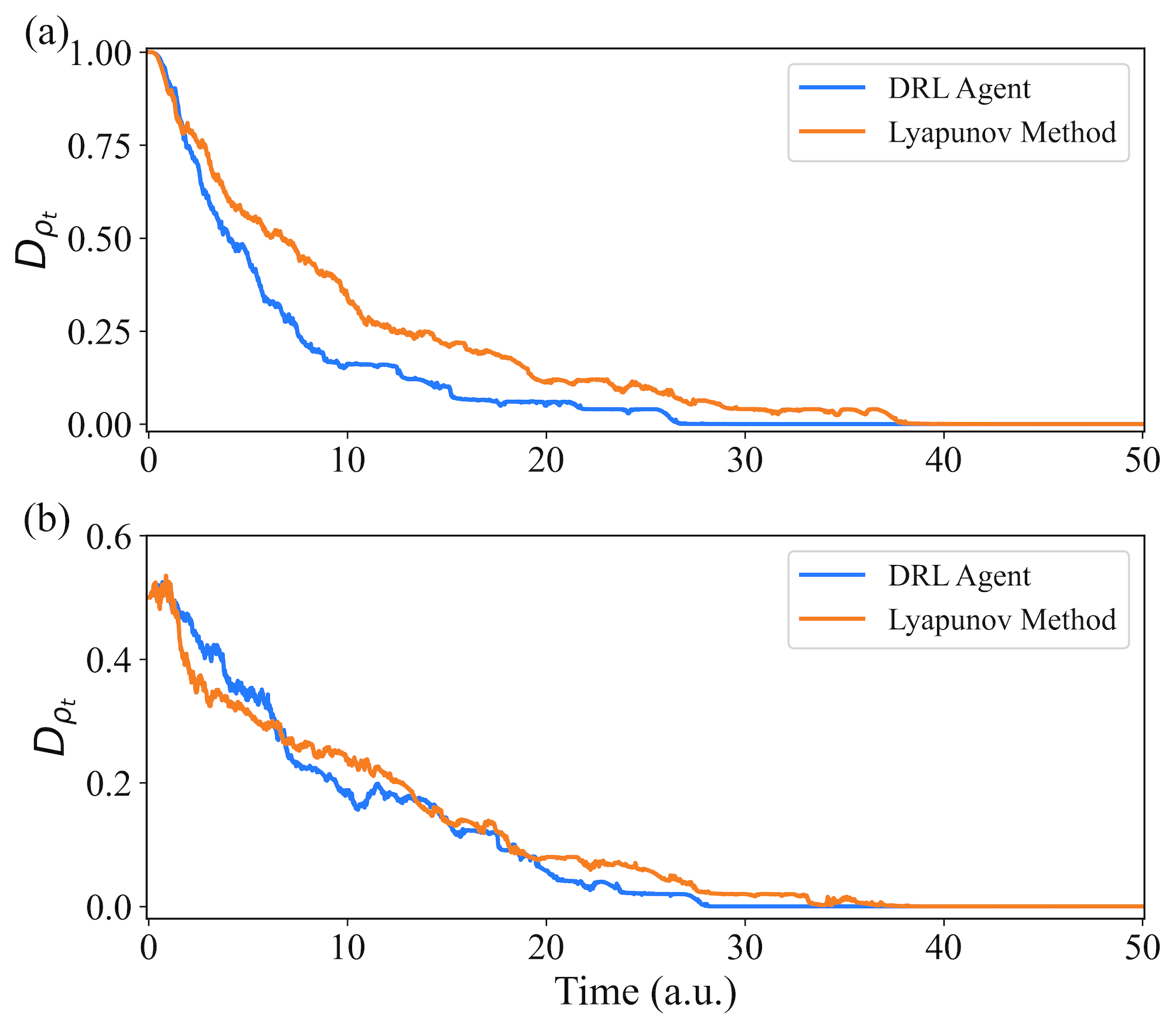}}
\caption{Evolution of the distance $D_{\rho_t}$ to stabilize two particular initial states (a) $\rho_{0}^1 = \text{diag}[0,1,0,0,0,0,0,0]$, (b) $\rho_{0}^2 = \text{diag}[1,0,0,0,0,0,0,0]$ to the target GHZ state under the control of the DRL agent (blue) and the Lyapunov method (orange). Each trajectory represents the average of 50 different stabilization processes for these specific initial states.}
\label{01_10}
\end{figure}

In addition, we also explore the evolution of two specific initial states, denoted as $\rho_{0}^1= \text{diag}[0,1,0,0,0,0,0,0]$ and $\rho_{0}^2= \text{diag}[1,0,0,0,0,0,0,0]$, mentioned in \cite{liu2016lyapunov} as examples. We repeat their stabilization $50$ times each to obtain averaged convergence curves that approximate the system's evolution. Fig.~\ref{01_10}(a) and Fig.~\ref{01_10}(b) depict the evolution of these two distinct initial states. The blue curve represents the evolution controlled by the DRL agent, while the orange curve represents the evolution controlled by the Lyapunov method from \cite{liu2016lyapunov}. It can be observed that the well-trained DRL agent not only achieves stable convergence to the target state but also showcases faster convergence compared to the Lyapunov method.

\begin{figure}[!t]
\centerline{\includegraphics[width=0.49\textwidth]{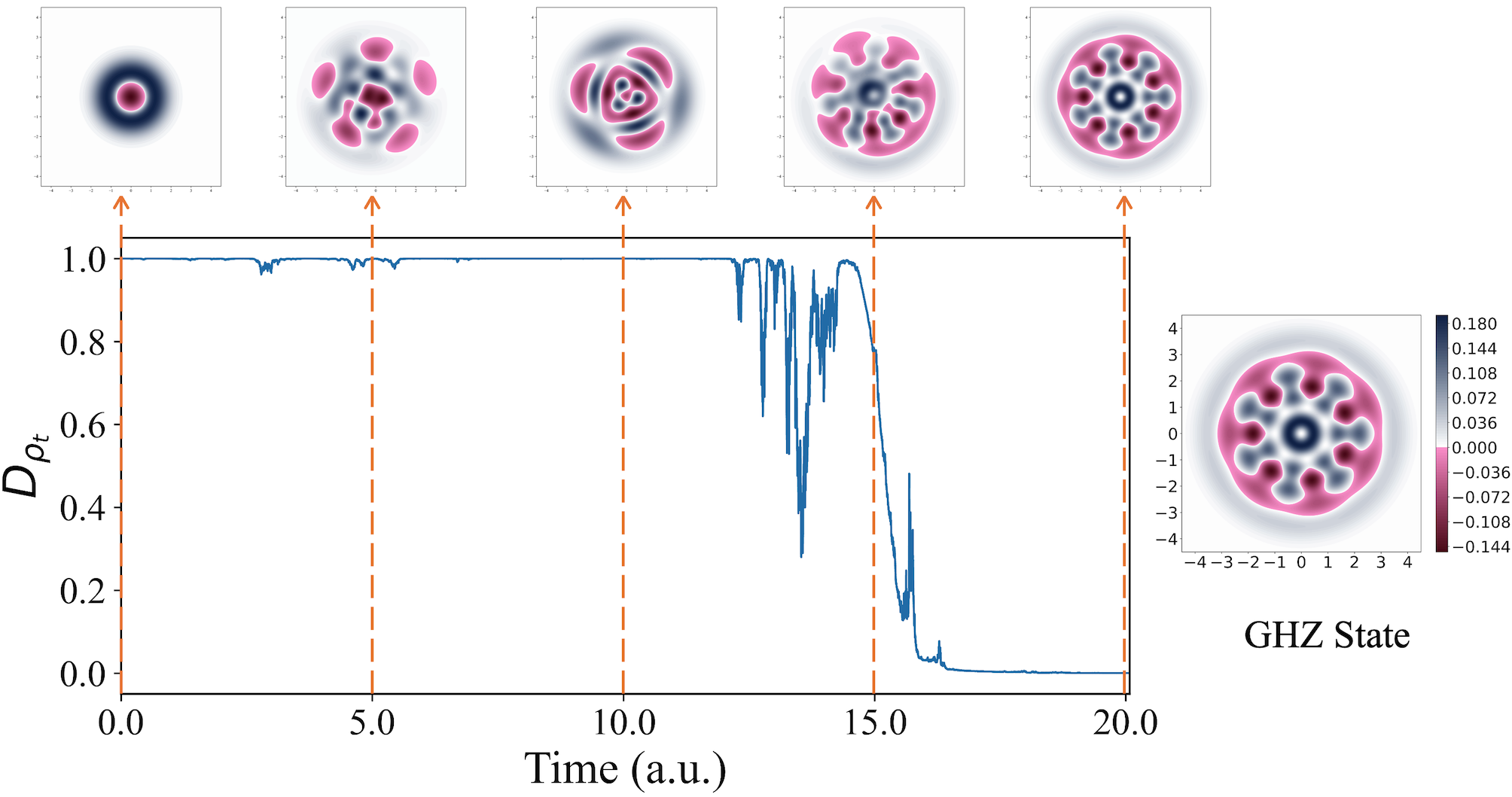}}
\caption{{One specific (not averaged)} evolutionary trajectory of the distance $D_{\rho_t}$ under the control of the DRL agent, starting from the initial state $\rho_{0}^1$. The top five images depict the Wigner function of the system state at different moments during the evolution. The figure on the right shows the phase-space probability distributions of the standard three-qubit GHZ state, which serves as a reference.}
\label{One_sample_winger}
\end{figure}

We randomly select a single trajectory under the control of a DRL agent with initial state $\rho_{0}^1$. The left subplot of Fig.~\ref{One_sample_winger} illustrates the evolutionary trajectory with $D_{\rho_t}$, and the top images display {the Wigner function of the system state} at five different evolution times. In contrast, the subplot on the right serves as a reference plot for the target three-qubit GHZ state. A comparison reveals that the phase-space distribution of the system state gradually approaches the target state over time, and at $t=20$ a.u., the system state is identical to the target state.

In practical agent training and application, uncertainties often exist. For example, the efficiency of measurements is typically not perfect, and there are frequently issues related to time delays in the feedback process. In the following two subsections we explore the robustness of our DRL agent to these two imperfections.

\subsubsection{Stabilization of the GHZ state with imperfect measurement}
\begin{figure}[!t]
\centerline{\includegraphics[width=0.49\textwidth]{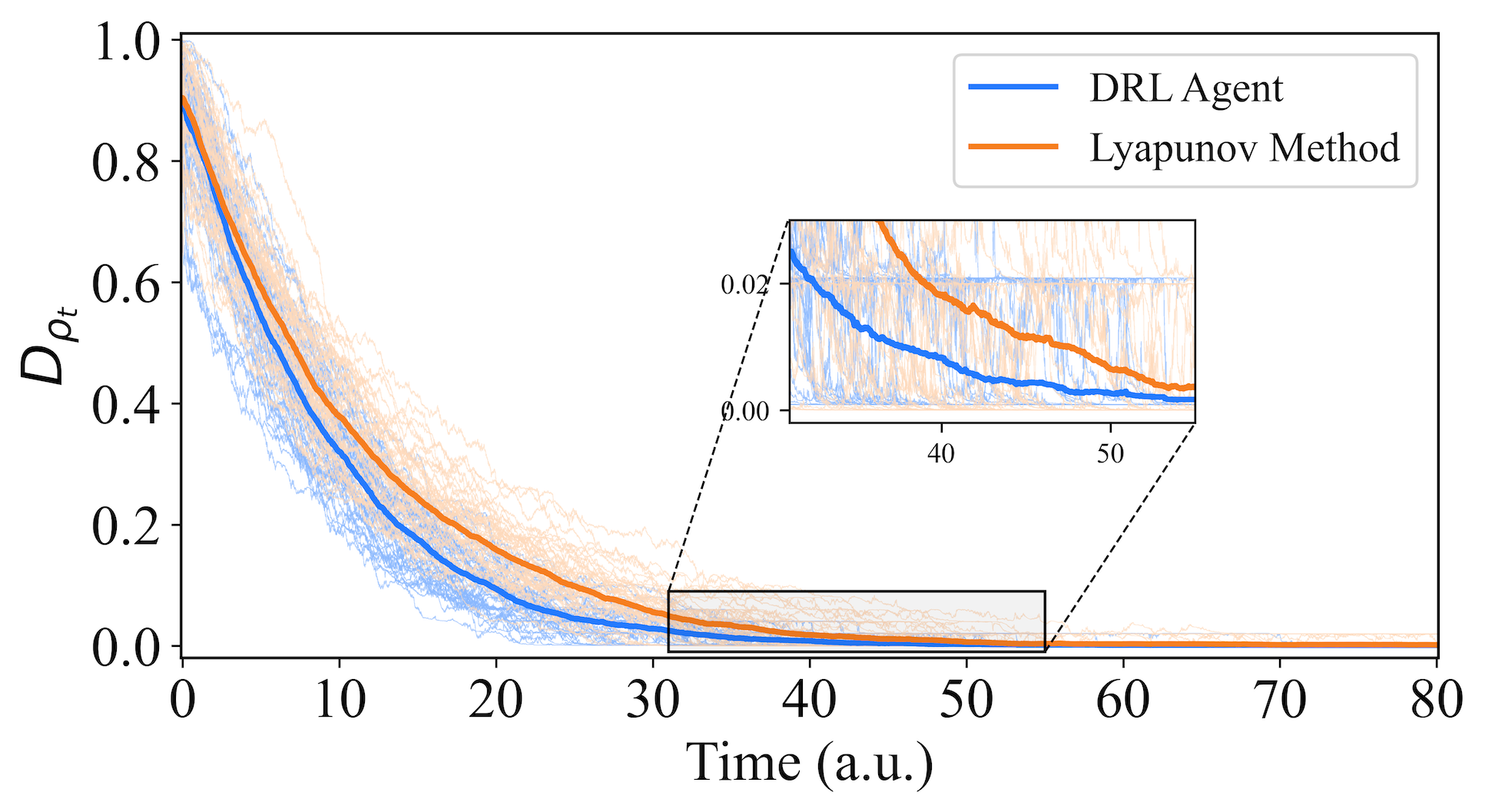}}
\caption{With a measurement efficiency $\eta_c=0.8$, under the control of the DRL agent and the Lyapunov-based method, $50$ random initial states stably evolve to the target GHZ state. The light lines represent the average evolution trajectories for different initial states, and the dark blue line represents the average trajectory across all different initial states.}
\label{pic:08efi}
\end{figure}

\noindent We first investigate the impact of reduced measurement efficiency, focusing on the robustness of an agent trained under the assumption of ``perfect case''. In this test, we consider a measurement efficiency of $\eta_c=0.8$, which represents a relatively high level achievable in current laboratory environments \cite{zhang2017quantum}. As shown in Fig.~\ref{pic:08efi}, {both the DRL-based agent and the Lyapunov-based method successfully stabilize $50$ randomly selected initial states to the target GHZ state. Moreover, the DRL agent demonstrates slightly superior performance. These results highlight that the DRL agent, trained under ideal conditions, retains significant robustness even in the presence of reduced measurement efficiency.}

\subsubsection{Stabilization of the GHZ state with time delay}
\begin{figure}[!t]
\centerline{\includegraphics[width=0.49\textwidth]{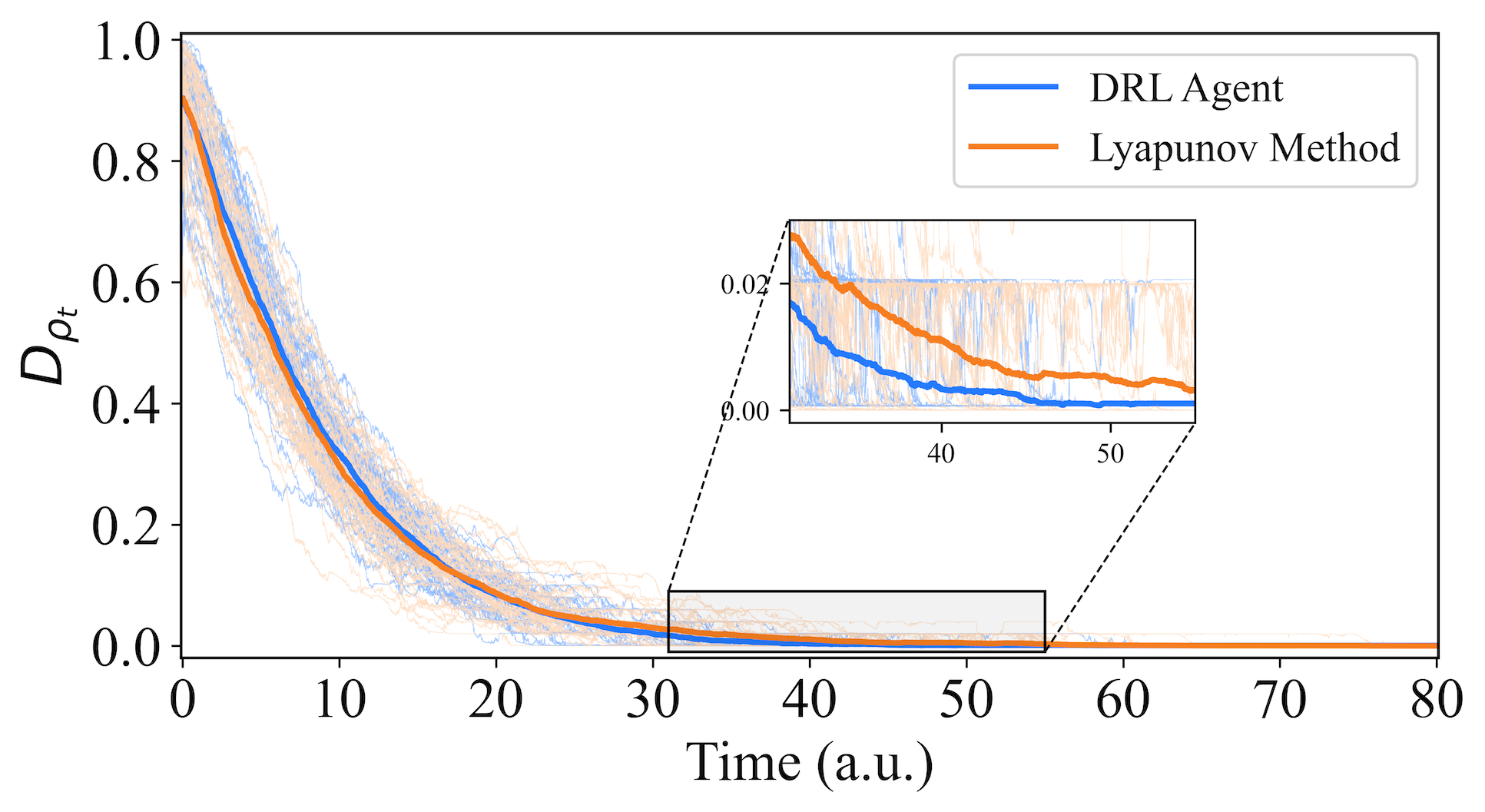}}
\caption{With a time delay $\mathcal{t}=0.05$ a.u., under the control of the DRL agent and the Lyapunov-based method, $50$ random initial states stably evolve to the target GHZ state. The light blue lines represent the average evolution trajectories for different initial states, and the dark blue line represents the average trajectory across all different initial states.}
\label{PicTimeDelay}
\end{figure}

\noindent We evaluate the performance of the trained agent under the presence of time delays in the feedback process. In rapidly evolving quantum systems, the time required for traditional computers to solve the SME \eqref{eq:SME} is often non-negligible. To account for this, we incorporate fixed time compensation during agent testing. For example, assuming a time delay of $\mathcal{t}=0.05$ a.u., the agent only receives the initial state $\rho_0$ as input from $t = 0$ a.u. to $t = 0.05$ a.u., and generates control signals based on this state to guide the system’s evolution. At $t = 0.051$ a.u., the agent receives the state $\rho_1$, and subsequently at each step, it processes the state $\rho_{t - 0.05}$, which corresponds to the state from $\mathcal{t} = 0.05$ a.u. earlier. As illustrated in Figure~\ref{PicTimeDelay}, using $50$ randomly selected initial states, {we observe that both the trained DRL agent and the Lyapunov-based method handle time delays effectively. Moreover, the DRL agent demonstrates superior performance compared to the Lyapunov-based approach.}

\section{Analyzing the Advantages of PNR Reward Function} \label{sec:CompRewards}
\noindent In this section, the 3-qubit example will be used to analyze the advantages of the PNR reward function. As described in Section~\ref{sec:myreward}, the PNR employs a partition parameter $d$ to divide the state space into two regions: the \textit{Proximity Zone} ($D_{\rho_t} < d$) and the \textit{Exploration Zone} ($D_{\rho_t} \geq d$). The reward function in each region follows the nonlinear form specified in \eqref{eq:rewardfunc}.

The parameters for the PNR are set as follows:
\begin{itemize}
\item $d = 0.001$, \quad $\mathfrak{e} = 2$, \quad $\mathfrak{f} = 10$;
\item $\overline{R}_P = 100$, \quad $\underline{R}_P = 1$, \quad $\overline{R}_E = 0$, \quad $\underline{R}_E = -0.1$.
\end{itemize}

To evaluate the effectiveness of the proposed approach, we consider several alternative reward functions for training DRL agent. For fair comparisons, all agents are trained with the same parameters for a given target state, with the only difference being the reward functions used during training. The purpose of the analysis is to give us a clear and comprehensive understanding to the impact of the various modules of the designed reward function on the control performance, as well as to compare its performance with existing reward function designs.

The following assumptions are made:
\begin{itemize}
\item The maximum evolution time of the system, denoted as $T_{\text{max}}$, is set to $100$ a.u.. For any trajectory where the system does not converge to the target state within $T_{\text{max}}$ (i.e., $D_{\rho_{T_{\text{max}}}} > 0.001$), it is considered a non-convergent trajectory. For the purpose of the follow-up calculation of the average time to convergence, the ``convergence time'' for these non-converging trajectories up to $100$ a.u. is set to $100$ a.u..
\item A metric termed the \textit{Stabilization Success Rate} is defined. During testing, $50$ distinct initial states are considered for stabilization to the target, with each initial state being stabilized under $50$ different environmental noise conditions. This results in a total of $50 \times 50 = 2500$ individual trajectories. The \textit{Stabilization Success Rate} is calculated as the ratio of the number of convergent trajectories within $T_{\text{max}} = 100$ to the total number of trajectories ($2500$).
\end{itemize}

\subsection{Reward Function Designs}
\noindent The reward functions tested are categorized as follows:
\begin{itemize}
    \item Partitioned Nonlinear Reward 1 (PNR1): This follows the structure of PNR, with the only difference being the parameters $\mathfrak{e} = 10$ and $\mathfrak{f} = 2$, such that the slope decreases as the distance diminishes.
    \item Partitioned Linear Reward (PLR): This design mirrors the partitioned structure of PNR but employs a linear reward function in each region instead of a nonlinear one. The bounds remain consistent with PNR.
    \item Partitioned Sparse Reward (PSR): \cite{bukov2018reinforcement, sivak2022model} use a sparse reward structure that provides rewards only at the last time step. We consider similar design ideas, retaining the partition structure for consistency:
    \begin{itemize}
        \item In the \textit{Proximity Zone}, the reward is a fixed positive constant.
        \item In the \textit{Exploration Zone}, no reward is assigned.
    \end{itemize}
    \item Non-Partitioned Nonlinear Reward (NPNR): This reward design does not use partitioning of the state space. Common reward function designs, such as those in \cite{porotti2022deep,perret2024preparation}, apply a uniform reward structure across all states, treating the entire state space equally without distinguishing between different regions. The NPNR reward function adopts the nonlinear form described in \eqref{eq:rewardfunc}, defined over the entire state space $D_{\rho_t} \in [0, 1]$. The rewards are non-positive, ranging from $[-1, 0]$, where the penalty magnitude decreases as $D_{\rho_t}$ gets closer to $0$.
    \item Non-Partitioned Linear Negative Reward (NPLNR): Similar to NPNR, this design does not partition the state space. The reward function is linearly defined over $D_{\rho_t} \in [0, 1]$ and is always non-positive $[-1, 0]$. The penalty decreases linearly as $D_{\rho_t}$ approaches $0$.
    \item Non-Partitioned Linear Positive Reward (NPLPR): The only distinction between this design and NPLNR is that the non-partitioned reward values are strictly positive, ranging from $[0, 100]$. The reward increases linearly as $D_{\rho_t}$ approaches $0$.
    \item Fidelity-Based Positive Reward (FPR): This non-partitioned reward is based on fidelity, a common approach in machine learning for quantum measurement feedback problems\cite{porotti2022deep,perret2024preparation}. Specifically, we adopt the reward function proposed in \cite{perret2024preparation}:
    \begin{equation}
        r = F(t)^4 + 4 \cdot F(t)^{25},
    \end{equation}
    where $F(t)$ is the fidelity at time $t$. The reward is positive and increases as $D_{\rho_t}$ approaches $0$.
\end{itemize}

\subsection{Performance Comparison}
\noindent To facilitate comparison, the reward functions are categorized into three types. Our PNR serves as the baseline for evaluating and comparing these categories.

\subsubsection{Category 1: Partitioned Nonlinear Rewards}
\begin{figure}[!t]
\centerline{\includegraphics[width=0.49\textwidth]{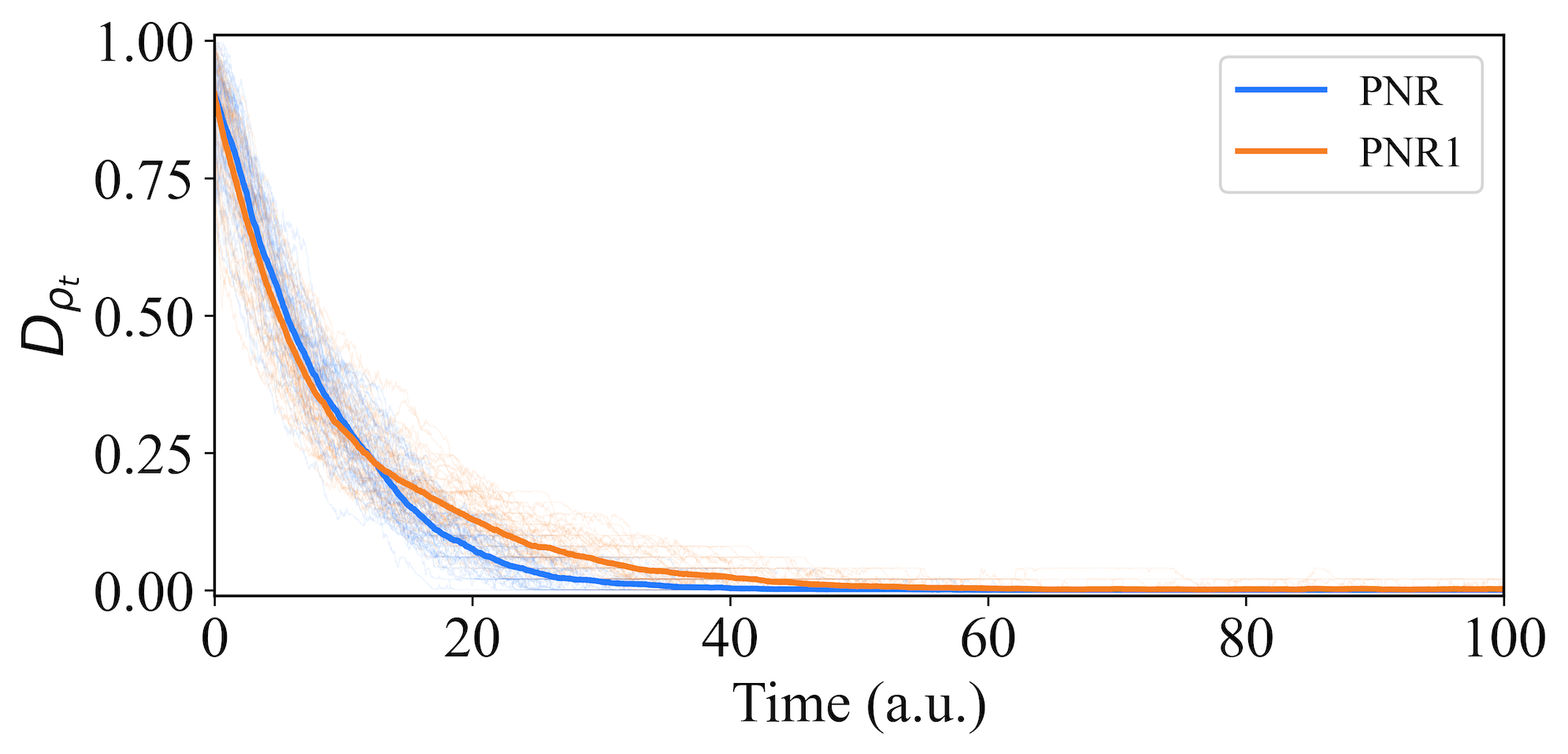}}
\caption{The effect of $\mathfrak{e}$ and $\mathfrak{f}$ parameters in the PNR reward function on the performance of the DRL algorithm. PNR: $\mathfrak{e}=2$ and $\mathfrak{f}=10$; PNR1:  $\mathfrak{e}=10$, $\mathfrak{f}=2$.}
\label{pic:CompRewards1}
\end{figure}

\noindent This category contains PNR1. The effect of the parameters $\mathfrak{e}$ and $\mathfrak{f}$ on the performance of the agent is evaluated. Using $50$ random initial states, each averaged over $50$ trajectories, the results in Fig.~\ref{pic:CompRewards1} show that PNR1 performs worse than PNR. The average convergence time for PNR1 is $12.19$ a.u., compared to $10.41$ a.u. for PNR. Additionally, the convergence success rate for PNR1 is $99.2\%$, lower than the $100\%$ success rate of PNR.  These results suggest that setting $\mathfrak{e}$ smaller than $\mathfrak{f}$ is more effective, meaning the reward function’s steepness increases as the target is approached.

\subsubsection{Category 2: Partitioned Linear and Sparse Rewards}
\begin{figure}[!t]
\centerline{\includegraphics[width=0.49\textwidth]{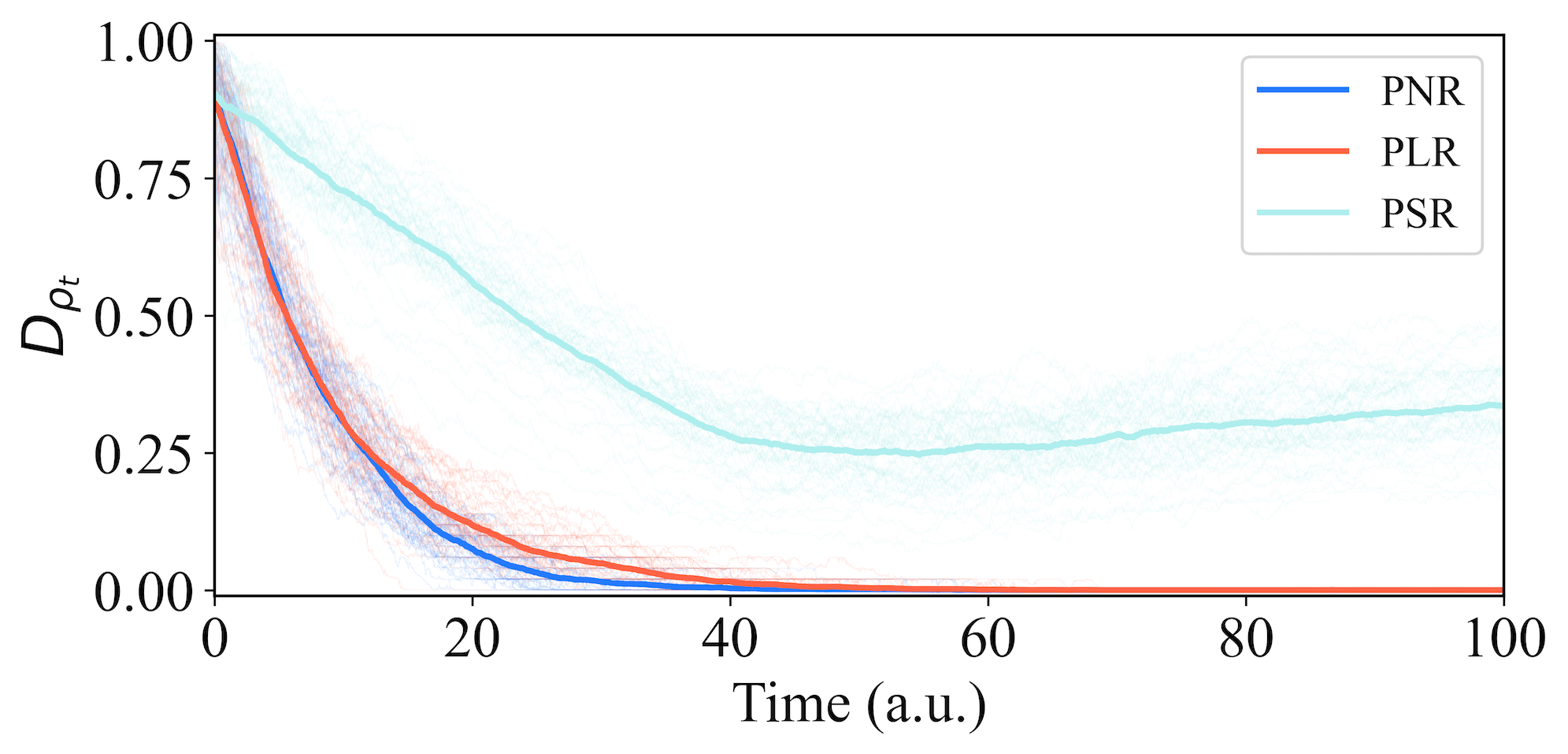}}
\caption{Effect of reward function on DRL algorithm under partitioning conditions.
PNR: partitioned nonlinear reward; PLR: partitioned linear reward; PSR: partitioned sparse reward.}
\label{pic:CompRewards2}
\end{figure}

\noindent This class includes PLR and PSR, whose only difference from our PNR is that the form of their reward functions are linear and sparse rewards, respectively. As shown in Fig.~\ref{pic:CompRewards2}, PLR achieves relatively good performance, with a convergence time of $11.80$ a.u. and a success rate of $99.76\%$, although it still underperforms compared to PNR. This highlights the superiority of our designed nonlinear reward function. In contrast, PSR performs poorly, as sparse rewards alone are insufficient for complex systems.

\subsubsection{Category 3: Non-Partitioned Rewards}
\begin{figure}[!t]
\centerline{\includegraphics[width=0.49\textwidth]{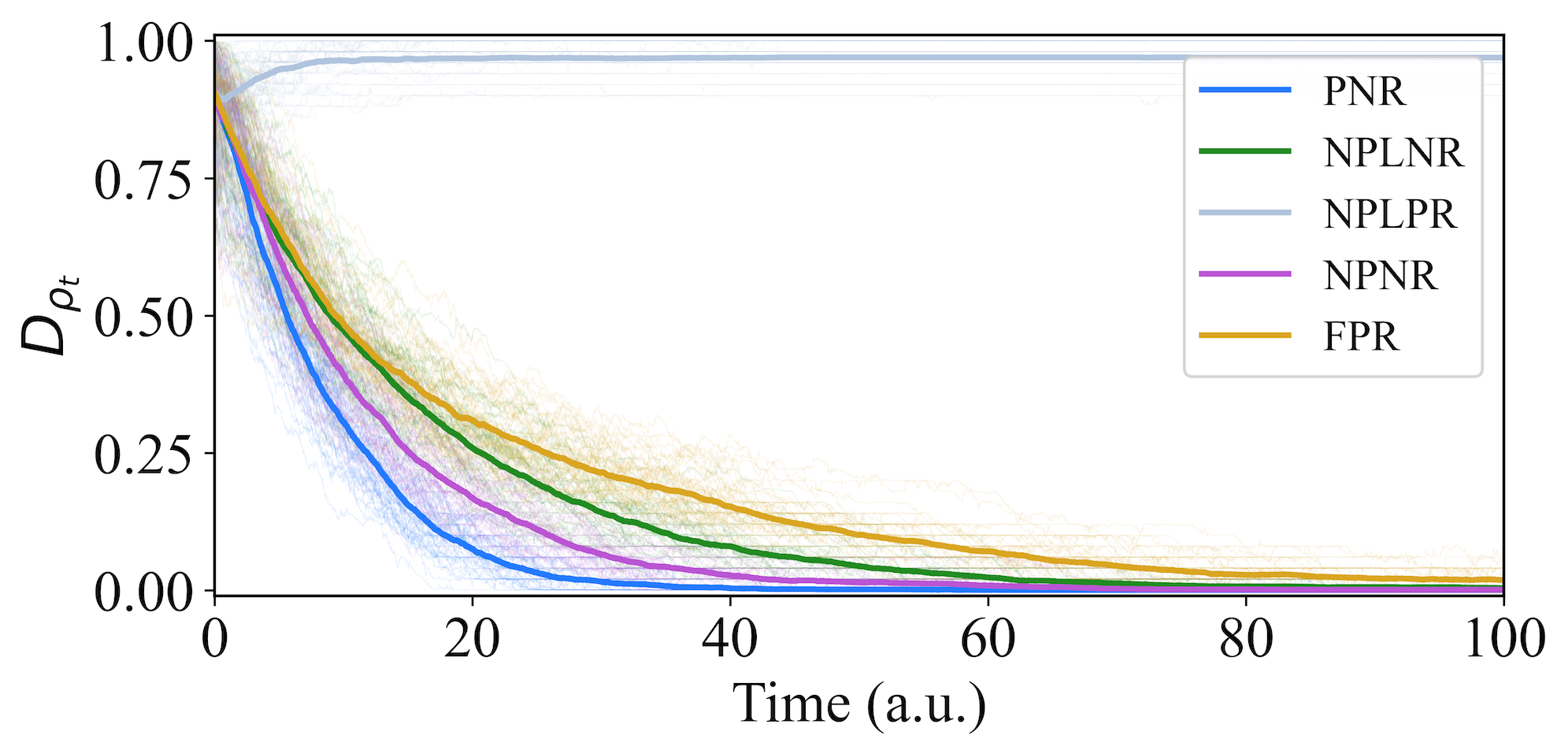}}
\caption{Effect of DRL algorithms with non-partitioned reward functions when not partitioned.
NPNR: nonlinear negative reward; NPLNR: linear negative reward; NPLPR: linear positive reward; FPR: fidelity-based reward.}
\label{pic:CompRewards3}
\end{figure}

\noindent This category evaluates NPNR, NPLNR, NPLPR, and FPR. These non-partitioned reward functions include linear, nonlinear, and fidelity-based, and the reward values are either positively encouraging (gradually increasing the positive reward value) or negatively spurring (gradually decreasing the negative reward value). Fig.~\ref{pic:CompRewards3} shows that among these designs, NPNR and NPLNR perform the best with convergence times of $13.60$ a.u. and $18.48$ a.u. and success rates of $99.88\%$ and $99.56\%$, respectively. These results suggest that decreasing negative penalties near the target is more effective than increasing positive rewards. In addition, the outperformance of NPNR over NPLNR highlights the role of nonlinear reward functions. The commonly used FPR performs poorly, with a convergence time and success rate of $43.59$ a.u. and $97.72\%$, respectively, highlighting its limitations in stabilizing complex quantum states. NPLPR's method of incrementing the positive rewards is nearly ineffective, which coincides with Remark $1$.

Our PNR outperforms other reward designs, demonstrating superior performance in stabilizing complex quantum states like the GHZ state. Nonlinear reward functions, especially when combined with partitioning, are shown to be more effective than linear or sparse reward structures in achieving faster convergence and higher success rates.

Table \ref{tab:reward_comparison} summarizes the parameters and characteristics of these reward function designs, as well as their performance.

\begin{table*}[htbp]
\centering
\caption{Comparison of Reward Functions}
\label{tab:reward_comparison}
\begin{tabular}{c|c|c|c|c|c|c|c|c}
\hline
\toprule
\textbf{Reward Type} & \textbf{Partitioned} & $\mathfrak{e}, \mathfrak{f}$ & \textbf{Reward Function} & \textbf{PZ Rewards} & \textbf{EZ Rewards} & \textbf{NP Rewards} & \textbf{Time (a.u.)} & \textbf{Success Rate} \\ \hline \hline
PNR & Yes & $\mathfrak{e}=2, \mathfrak{f}=10$ & Nonlinear  & [1, 100] & [-0.1, 0] & \textbackslash & 10.41 & 100\% \\ \hline
PNR1 & Yes & $\mathfrak{e}=10, \mathfrak{f}=2$ & Nonlinear  & [1, 100] & [-0.1, 0] & \textbackslash & 12.19 & 99.2\% \\ \hline
PLR & Yes & \textbackslash & Linear  & [1, 100] & [-0.1, 0] & \textbackslash & 11.8 & 99.76\% \\ \hline
PSR & Yes & \textbackslash & Sparse  & 1 & 0 & \textbackslash & 53.25 & 62.08\% \\ \hline
NPNR & No & \textbackslash & Nonlinear  & \textbackslash & \textbackslash & [-1, 0] & 13.6 & 99.88\% \\ \hline
NPLNR & No & \textbackslash & Linear  & \textbackslash & \textbackslash & [-1, 0] & 18.48 & 99.56\% \\ \hline
NPLPR & No & \textbackslash & Linear  & \textbackslash & \textbackslash & [0, 100] & 97.23 & 3.08\% \\ \hline
FPR & No & \textbackslash & Fidelity-Based  & \textbackslash & \textbackslash & Fidelity & 43.59 & 97.72\% \\ \hline \hline
\multicolumn{9}{c}{\textbf{PZ (EZ) Rewards}: Reward values in Proximity (Exploration) Zone; \quad \textbf{NP Rewards}: Reward values when not partitioned.}\\
 \bottomrule
\end{tabular}
\end{table*}

\section{Conclusions}
\label{Conclusion}

\noindent In this work, we designed a DRL agent and applied it to measurement-based feedback control for stabilizing quantum entangled states. With a designed reward function, the trained agent can quickly stabilize the quantum system to the target state with high fidelity and demonstrates strong robustness.

The DRL agent was tested under a ``perfect case'' and its performance was 
mainly compared to a Lyapunov-based switching method, and several other DRL algorithms with different reward functions. The results showed that our approach achieve comparable fidelity in a shorter time, potentially mitigating noise caused by prolonged interactions. The agent was also evaluated under ``imperfect cases,'' including low measurement efficiency and feedback delay, where it maintained great performance under these challenging conditions.

To analyze our method, we conducted ablation studies on the reward function to identify the role of each component in achieving the final performance. We also compared our reward function with fidelity-based designs from the literature and several variants of our reward function, showing the superior performance of our approach.

The proposed DRL-based framework has the potential to be extended to stabilize any multi-qubit system. Naturally, it is expected that as the system dimension increases, both the training time and the time required for state estimation using the SME will grow. Future research could explore methods to reduce the dimensionality of the state space and enhance the speed of quantum state estimation based on the SME. Furthermore, developing control strategies that rely on less system information, such as using only measurement currents, may allow for efficient stabilization of the target state with sufficient fidelity. These directions aim to further improve the practical applicability of the proposed control framework.

{\appendix[PPO algorithm]
\noindent In the field of DRL, PPO algorithm has become a prominent method due to its effectiveness and stability in policy optimization. This appendix is dedicated to elucidating the core algorithmic concepts of PPO and providing a detailed description of its mechanisms. The aim is to provide the reader with a comprehensive understanding of how the algorithm works and facilitate its application in practical scenarios.The core of PPO is to maximize the expected cumulative returns by optimizing the strategy parameters while ensuring that the updates remain stable. The following paragraphs delve into the specifics of these algorithmic ideas.

\subsection{Core Algorithmic Ideas of PPO}
\noindent The probability of each sequence occurring is multiplied by its corresponding cumulative reward, and the sum of these products yields the expected reward. The probability of a specific $\tau$ sequence occurring given $\theta$ is defined as:
\begin{equation}
    p_{\theta}(\tau) = \prod_{t=0}^{m}p_{\theta}(a_{t}|\rho_{t})p(\rho_{t+1}|\rho_{t},a_{t}).
\end{equation}
We denote $p(\text{EVENT})$ to signify the probability of the occurrence of the EVENT. For example, $p_{\theta}(a_t|\rho_t)$ represents the probability of agent to choose action $a_t$ given $\rho_t$ while $p(\rho_{t+1}|\rho_{t},a_{t})$ represents the probability of environment transiting at $\rho_{t+1}$ from $\rho_t$ given the action $a_t$ applied.

When the parameter $\theta$ is given, the expected value of the total reward, denoted as $J(\theta)$, is evaluated as the weighted sum of each sampled $\tau$ sequence, expressed by
\begin{equation}
\label{J_theta}
    J(\theta) = \sum_{\tau}R(\tau)p_{\theta}(\tau) := \underset{\tau\sim p_{\theta}(\tau)}{\mathbb{E}}[R(\tau)].
\end{equation}
To maximize $J(\theta)$, which indicates that our chosen policy parameters $\theta$ can lead to higher average rewards, we adopt the well-known gradient descent method. Thus, we take the derivative of the expected reward $J(\theta)$ in \eqref{J_theta}, resulting in the expression shown in 
\begin{alignat}{1}
        \nabla J(\theta)& = \sum_{\tau} R(\tau) \nabla p_{\theta}(\tau) \nonumber \\
                        & = \sum_{\tau} R(\tau) p_{\theta}(\tau) \nabla \log \, p_{\theta}(\tau)  \nonumber \\
                        & = \sum_{\tau} R(\tau) p_{\theta}(\tau)\sum_{t=0}^{m} \nabla \log \,p_{\theta}(a_{t} | \rho_{t}) \label{PG} \\
                        & = \underset{\tau\sim p_{\theta}(\tau)}{\mathbb{E}} \Bigg[R(\tau)\sum_{t=0}^{m} \nabla \log \,p_{\theta}(a_{t} | \rho_{t})\Bigg] \nonumber\\
                        & \approx \underset{(\rho_t,a_t)\sim \pi_{\theta}}{\mathbb{E}} \Bigg[\nabla \log \,p_{\theta}(a_{t} | \rho_{t}) A^{\theta}(\rho_t, a_t)\Bigg].\nonumber
\end{alignat}
We use $\nabla f(x)=f(x)\nabla \log f(x)$ to derive the second row of \eqref{PG}. The last approximate equation is a result of practical gradient computations, where instead of calculating the expected reward for an entire trajectory, rewards contributed by each individual state-action pair ($\rho, a$) are computed separately. These individual rewards are then summed up to obtain the total cumulative reward for the optimization process. The direction of the policy update $\pi_\theta(a|\rho)$ is biased towards favoring state-action pairs that contribute to higher cumulative rewards within the sequence. For instance, if an action $a$ executed in state $\rho$ leads to a positive cumulative discounted reward, the subsequent update will enhance the probability of choosing action $a$ in state $\rho$, while diminishing the likelihood of selecting other actions. The update equation for the parameters $\theta$ is as follows: 
\begin{equation}
    \theta = \theta + \eta \nabla J(\theta),
\end{equation}
where $\eta$ is the learning rate.

Once the policy $\pi_\theta(a|\rho)$ is updated, it necessitates the reacquisition of training data prior to the subsequent policy update. This arises due to the alteration in the probability distribution $p_{\theta}(\tau)$ brought about by the modified policy. Following data sampling, the parameter $\theta$ undergoes refinement, leading to the discarding of all prior data. Subsequent parameter updates mandate the collection of fresh data, constituting the fundamental principle underlying the conventional PG algorithm. However, in the context of quantum systems, the process of sampling system information is often characterized by time-intensive and computationally demanding operations. For instance, after each measurement, a classical computer is requisitioned to solve the SME \eqref{eq:SME} to ascertain the system's state in the subsequent moment. This inability to reutilize previously acquired data contributes to a protracted training process. To address this challenge, an additional strategy $\pi_{\theta^{\prime}}$ is introduced, mirroring the architecture of $\pi_{\theta}(a|\rho)$. Instead of directly engaging with the environment for data gathering, the primary agent $\pi_{\theta}$ employs the auxiliary agent $\pi_{\theta^{\prime}}$ to interact with the environment and accumulate data. The objective is to subsequently utilize this data to train $\pi_{\theta}$ multiple times, effectively reducing the computational and resource demands for data collection. Ensuring the consistency of data sampled by $\pi_{\theta^{\prime}}$ with that of $\pi_{\theta}$, importance sampling \cite{xie2019towards} is introduced to facilitate this synchronization process. This approach contributes to enhancing data reuse and the overall efficiency of the training procedure. \eqref{PG} is updated as:
\begin{alignat}{1}
    & \nabla J(\theta) \nonumber \\
    & = \underset{(\rho_t,a_t)\sim \pi_{\theta^\prime}}{\mathbb{E}} \Bigg[\frac{p_{\theta}(\rho_t,a_t)}{p_{\theta^{\prime}}(\rho_t,a_t)} \nabla \log p_{\theta}(a_{t} |\rho_{t}) A^{\theta^{\prime}}(\rho_t, a_t)\Bigg] \nonumber \\ 
    & = \underset{(\rho_t,a_t)\sim \pi_{\theta^\prime}}{\mathbb{E}} \Bigg[\frac{p_{\theta}(a_t | \rho_t)p_{\theta}(\rho_t)}{p_{\theta^{\prime}}(a_t|\rho_t)p_{\theta^{\prime}}(\rho_t)} \nabla \log p_{\theta}(a_{t} |\rho_{t}) A^{\theta^{\prime}}(\rho_t, a_t) \Bigg] \nonumber \\
    & =\underset{(\rho_t,a_t)\sim \pi_{\theta^\prime}}{\mathbb{E}} \Bigg[\frac{p_{\theta}(a_t | \rho_t)}{p_{\theta^{\prime}}(a_t|\rho_t)} \nabla \log p_{\theta}(a_{t} |\rho_{t}) A^{\theta^{\prime}}(\rho_t, a_t) \Bigg]. \label{AD gradient}
\end{alignat}
Here, all the state-action pairs $(\rho_t,a_t)$ (or alternatively, all trajectories $\tau \in \mathbb{D}$) are sampled from $\pi_{\theta^{\prime}}$. The term ${p_{\theta}(\rho_t,a_t)} / {p_{\theta^{\prime}}(\rho_t,a_t)}$ represents the importance weight, which dynamically adjusts the weighting of data sampled by $\pi_{\theta^{\prime}}$ in real time to more accurately estimate the expected value under the target policy $\pi_{\theta}$.

The corresponding objective function from \eqref{AD gradient} can be calculated as:
\begin{equation}
    J^{\theta^{\prime}}(\theta)=\underset{(\rho_t,a_t)\sim \pi_{\theta^\prime}}{\mathbb{E}} \Big[\frac{p_{\theta}(a_t | \rho_t)}{p_{\theta^{\prime}}(a_t|\rho_t)} A^{\theta^{\prime}}(\rho_t, a_t)\Big].
    \label{AD objective}
\end{equation}

Nonetheless, in the absence of constraint, such as when $A^{\theta^{\prime}}(\rho_t, a_t)>0$, indicating the desirability of specific action-state combinations, the agent's inclination would be to elevate their likelihood, effectively amplifying the ${p_{\theta}(a_t | \rho_t)}/{p_{\theta^{\prime}}(a_t|\rho_t)}$ value. This scenario can lead to policy learning inaccuracies and an erratic learning process, impeding convergence. To counteract this, the PPO introduces a pivotal mechanism, termed the ``clip ratio". This clip ratio imposition serves to confine the proportions between the new and preceding policies, thereby ensuring congruence and augmenting the algorithm's dependability. The following equation demonstrates the PPO Clipping algorithm, incorporating the clipping term to bound the difference between $p_{\theta}$ and $p_{\theta^{\prime}}$ during the policy update. 
\begin{alignat}{1}
    & J_{\rm{PPO}}^{\theta^{\prime}}(\theta) \nonumber \\ 
    & \approx \underset{(\rho_t,a_t)\sim \pi_{\theta^\prime}}{\mathbb{E}} \min \Bigg(\varrho A^{\theta^{\prime}}(\rho_t, a_t), \label{PPO} \\
    & \qquad \qquad \qquad \qquad {\rm{clip}} \Bigg(\varrho, 1-\varsigma, 1+\varsigma\Bigg)A^{\theta^{\prime}}(\rho_t, a_t)\Bigg), \nonumber
\end{alignat}
where $\varrho = {p_{\theta}(a_t|\rho_t)}/{p_{\theta^{\prime}}(a_t|\rho_t)}$. The last two terms $1-\varsigma$, and $1+\varsigma$ in the clip function limit the boundaries of the first term. $\varsigma$ is a hyperparameter, typically set to $0.1$ or $0.2$. Exhaustively considering all possible sequences is typically infeasible, and thus in practical training, the objective function \eqref{PPO} is often formulated in the following manner:
\begin{alignat}{1}
    & J_{\rm{PPO}}^{\theta^{\prime}}(\theta) \nonumber \\ 
    & \approx \frac{1}{Nm} \sum_{\tau \in \mathbb{D}} \sum_{t=0}^{m} \min \Bigg(\varrho A^{\theta^{\prime}}(\rho_t, a_t), \label{PPO_Final} \\
    & \qquad \qquad \qquad \qquad \quad {\rm{clip}} \Bigg(\varrho, 1-\varsigma, 1+\varsigma\Bigg)A^{\theta^{\prime}}(\rho_t, a_t)\Bigg), \nonumber
\end{alignat}
where $N$ and $m$ represent finite real numbers that respectively signify the count of collected sequences and the maximum number of steps within each sequence.

Based on the above, we can see that PPO has three sets of network parameters for its update strategy:
\begin{itemize}
    \item One set of main policy parameters $\theta$, which is updated every time.
    \item One set of policy parameter copies $\theta^\prime$, which interact with the environment and collect data. They utilize importance sampling to assist in updating the main policy parameters $\theta$. Typically, $\theta^\prime$ is updated only after several updates of $\theta$ has been performed.
    \item One set of value network parameters $\phi$, which are updated based on the collected data using supervised learning to update the evaluation of states. They are also updated every time.
\end{itemize}

\subsection{Generalized parameter selection for training agents}
\noindent {\bf{Neural Network Architecture:}} Each policy $\pi_\theta$ is represented by a neural network that maps a given state $\rho$ to a probability distribution over actions $a$. The action distribution is modeled as a gaussian distribution. The input layer is processed by two fully connected hidden layers, each with $128$ neurons, accompanied by a linear output layer with the same dimension as the action space ($2$ in this paper). All hidden layers use the Tanh activation function. The value function $V^{\phi}(\rho_t)$ is composed of a similar neural network architecture, with the only difference being that the output layer is a single linear unit used to estimate the state-value function. The value function $V^{\phi}(\rho_t)$ is estimated using the temporal difference (TD) method \cite{sutton2018reinforcement}. Then, the generalized advantage estimator (GAE) \cite{schulman2015high} is employed to compute the advantage function in \eqref{R_tau}, which is subsequently used in \eqref{PPO} to calculate the gradient for updating the policy $\pi_\theta$. 

\noindent {\bf{Learning Rate:}} The learning rate is a hyperparameter that determines the step size of the algorithm's updates based on observed rewards and experiences during training. In our training process, the learning rate $\eta$ is not a constant value but follows a linear schedule. Our learning rate starts at $\eta=5*10^{-7}$ and linearly decreases over time during the training process. This allows the algorithm to explore more in the early stages of training when the policy might be far from optimal. As the training progresses, the policy approaches convergence, and the learning rate decreases to promote stability in the learning process and fine-tune the policy around the optimal solution. This helps the DRL agent achieve better performance and stability during the training process. Please refer to \cite{sutton2018reinforcement,peng2018deepmimic} for more details.}

\bibliographystyle{IEEEtran}
\bibliography{IEEEabrv,ref}

\vspace{12pt}

\end{document}